\documentclass[epsfig,12pt]{article}
\usepackage{epsfig,amssymb,euscript}
\usepackage{amsmath,amscd}

\numberwithin{equation}{section}
\newcommand{\be}{\begin{eqnarray}}
\newcommand{\ee}{\end{eqnarray}}
\newcommand{\bea}{\begin{eqnarray}}
\newcommand{\eea}{\end{eqnarray}}
\newcommand{\ba}{\begin{array}}
\newcommand{\ea}{\end{array}}

\newcommand{\nn}{\nonumber \\}

\newcommand{\p}[1]{(\ref{#1})}

\DeclareMathOperator{\SU}{\mathit{SU}}

\DeclareMathOperator{\Spin}{\mathit{Spin}}

\DeclareMathOperator{\Cliff}{Cliff}

\newcommand{\SE}{\mathrm{SE}}

\newcommand{\id}{\mathbf{1}}
\newcommand{\trsp}{\mathrm{T}}

\DeclareMathOperator{\diag}{diag}

\DeclareMathOperator{\re}{Re} 
\DeclareMathOperator{\im}{Im}

\DeclareMathOperator{\Ric}{Ric}

\DeclareMathOperator{\vol}{vol}

\newcommand{\ii}{\mathrm{i}}
\newcommand{\dd}{\mathrm{d}}
\newcommand{\diff}{\mathrm{d}}
\newcommand{\me}{\mathrm{e}}

\newcommand{\de}{\partial}

\newcommand{\bR}{\mathbb{R}}
\newcommand{\bC}{\mathbb{C}}

\newcommand{\bCP}{\mathbb{C}P}

   \def\b{\beta}   \def\l{\lambda}  
         
\def\e{\epsilon}

\begin{document}



\begin{titlepage}

\vfill

\begin{flushright}
Imperial/TP/041105\\
CERN-PH-TH/2005-184\\
HUTP-04/A0051\\
hep-th/0510125\\
\end{flushright}

\vfill

\begin{center}
   \baselineskip=16pt
   {\Large\bf Supersymmetric $AdS_5$ Solutions \\[5mm] of Type IIB Supergravity}
   \vskip 2cm
      Jerome P. Gauntlett$^{1}$, Dario Martelli$^{2}$, James Sparks$^{3}$ and Daniel Waldram$^{1}$
   \vskip .6cm
   \begin{small}
      \textit{$^{1}$Blackett Laboratory, Imperial College\\
        London, SW7 2AZ, U.K.}
        \end{small}
 \vskip .6cm
   \begin{small}
      \textit{$^{2}$Department of Physics, CERN Theory Division\\
        1211 Geneva 23, Switzerland}
        \end{small}
\vskip .6cm
   \begin{small}
      \textit{$^{3}$Department of Mathematics, Harvard University\\
        One Oxford Street, Cambridge, MA 02138, U.S.A.\\
    {\it and}\\
    Jefferson Physical Laboratory, Harvard University\\
    Cambridge, MA 02138, U.S.A.}
   \end{small}
\end{center}

\vfill

\begin{center}
\textbf{Abstract}
\end{center}
We analyse the most general bosonic supersymmetric solutions of
type IIB supergravity whose metrics are warped products of
five-dimensional anti-de Sitter space ($AdS_5$) with a
five-dimensional Riemannian manifold $M_5$. All fluxes are allowed
to be non-vanishing consistent with $SO(4,2)$ symmetry. We show
that the necessary and sufficient conditions can be phrased in
terms of a local identity structure on $M_5$. For a special class,
with constant dilaton and vanishing axion, we reduce the problem
to solving a second order non-linear ODE. We find an
exact solution of the ODE which reproduces a solution first found
by Pilch and Warner. A numerical analysis of the ODE reveals an
additional class of local solutions.

\begin{quote}

\end{quote}

\vfill

\end{titlepage}

\section{Introduction}

The AdS/CFT correspondence \cite{mal} is one of the most important
developments in string theory. It is therefore an important issue
to understand the geometric structures underpinning the
correspondence. On the one hand such an understanding can lead to
new explicit examples where one can make detailed comparisons with
the dual field theory and which can also suggest further
generalisations. On the other hand, and more generally, a precise
statement of the underlying geometry is the foundation for
progress without recourse to explicit examples. By analogy, recall
that our understanding of Calabi--Yau geometry has been made
without a single non-trivial explicit compact Calabi--Yau 3-fold
metric having been constructed.

In ref. \cite{gmsw} we analysed the most general kind of solutions
of $D=11$ supergravity that can be dual to a four-dimensional
superconformal field theory. These bosonic supersymmetric
solutions have a metric that is a warped product of $AdS_5$ with a
six-dimensional Riemannian manifold $M_6$. In order that the
$SO(4,2)$ isometry group of $AdS_5$ is a symmetry group of the
full solution, the four-form field strength has non-vanishing
components only on $M_6$.  We used the, by now, standard technique
of analysing the canonical $G$-structure dictated by supersymmetry
\cite{Gauntlett:2002sc,Gauntlett:2002nw,Gauntlett:2002fz} in order
to obtain necessary and sufficient conditions for supersymmetry.
We showed that the geometry on $M_6$ admits a local
$SU(2)$-structure and that this implies that $M_6$ is determined,
in part, by a one parameter family of K\"ahler metrics.

We further analysed a special sub-class of solutions by imposing
the condition that $M_6$ is complex and we used the results to
construct several new classes of compact examples of $M_6$ in
explicit form. We showed that one sub-class of solutions leads to
new type IIA and type IIB solutions with $AdS_5$ factors, via
dimensional reduction and T-duality, respectively. In particular,
the type IIB solutions turn out to be direct product backgrounds
$AdS_5\times X_5$ with $X_5$ a Sasaki--Einstein manifold and only
the self-dual five-form non-vanishing and proportional to the sum
of the volume forms on $AdS_5$ and $X_5$ -- see
\cite{Klebanov:1998hh,Figueroa-O'Farrill:1998nb,Acharya:1998db,Morrison:1998cs}
for a general discussion of such backgrounds. This is an
interesting class of solutions since the dual SCFT can be
identified as that arising on a stack of D3-branes transverse to
the Calabi--Yau three-fold cone based on $X_5$. The most
well-known examples of five-dimensional Sasaki--Einstein
manifolds, and until recently the only explicit examples, are
$S^5$ and $T^{1,1}$; the corresponding IIB solutions are dual to
$N=4$ super Yang--Mills theory and an $N=1$ superconformal field
theory discussed in \cite{Klebanov:1998hh, Morrison:1998cs},
respectively. The solutions found in \cite{gmsw} led to an
infinite number of new explicit Sasaki--Einstein metrics on
$S^2\times S^3$ called $Y^{p,q}$ \cite{gmsw2}. The dual conformal
field theories have now been identified \cite{DJ,BBC,quivers} and
there have been many further checks and developments. The
$Y^{p,q}$ metrics were generalised to all dimensions in
\cite{gmsw3} and were recently further generalised  to
the $L^{a,b,c}$ metrics in \cite{Cvetic:2005ft,Cvetic:2005vk} (see
also \cite{Martelli:2005wy}).

The analysis of \cite{gmsw} covered $AdS_5$ geometries in $D=11$
supergravity preserving $N=1$ supersymmetry. A refinement of this
analysis was recently carried out in \cite{Lin:2004nb}, where the
additional conditions imposed by $N=2$ supersymmetry were studied.

In this paper we will generalise the M-theory analysis of
\cite{gmsw} to type IIB string theory. In particular, we go beyond
the Sasaki--Einstein class and analyse the most general bosonic
supersymmetric solutions of type IIB supergravity with a metric
that is a warped product of $AdS_5$ with $M_5$. In addition we
allow all of the NS-NS and R-R bosonic fields to be non-vanishing
consistent with $SO(4,2)$ symmetry. Once again, following
\cite{Gauntlett:2002sc,Gauntlett:2002nw,Gauntlett:2002fz}, we
analyse the $G$-structure defined by the Killing spinors. We find
that the most general geometries have a  local identity
structure, or equivalently a canonically defined frame, and we use
this to determine the necessary and sufficient conditions for
supersymmetry. The geometries have a canonically defined Killing
vector, which corresponds to the $U(1)$ R-symmetry of the dual
SCFT. We also show that for these solutions supersymmetry implies
the equations of motion, just as we saw in \cite{gmsw}.

To construct explicit solutions we further restrict our
considerations to the special case of constant dilaton and
vanishing axion with some additional restrictions imposed on the
geometry. We can then reduce the entire problem to solving a
second order non-linear ODE. We find one solution in closed form,
which turns out to be a solution first obtained by Pilch--Warner
\cite{pw}  (constructed by uplifting a solution first
found in five-dimensional gauged supergravity \cite{kpw}). This
solution has constant dilaton and vanishing axion, but
non-vanishing three-forms and self-dual five-form; it has been
identified \cite{Karch:1999pv,Freedman:1999gp} as being dual to an
$N=1$ supersymmetric fixed point discovered by Leigh and Strassler
\cite{Leigh:1995ep}. A numerical analysis of our ODE leads to a
continuous family of local solutions. We
show that they lead to complete metrics on $S^5$, but a detailed
analysis indicates that neither the three-form fluxes nor the
spinors are globally defined. It is not clear to us whether or not
these solutions can be given a physical interpretation. It is also
possible that other solutions of the ODE lead to interesting
solutions, but we leave this for future work.

The plan of the rest of paper is as follows. In section 2 we
outline our conventions for type IIB supergravity. Section 3
derives the necessary and sufficient conditions for the most
general supersymmetric solutions with $AdS_5$ factors. For the
convenience of the reader, we have summarised the main results, in
a somewhat self contained way, in section 3.6. Section 4 continues
the analysis by introducing local coordinates. The discussion of the special
class of solutions, including vanishing axion and constant dilaton, and the recovery
of the Pilch--Warner solution, is presented in section 5.
Section 6 briefly concludes. We have relegated some technical material to several appendices.

\section{Type IIB equations and conventions}
\label{conv}

We begin by presenting the equations of motion and supersymmetry
transformations for bosonic configurations of type IIB
supergravity \cite{Schwarz:qr,Howe:sr} in the conventions given in
appendix~\ref{spinors}. Essentially we are following \cite{Schwarz:qr},
with some minor changes, including the signature of the metric.

The conditions for a bosonic geometry to preserve some
supersymmetry are
\begin{equation}
\label{10d-susy}
\begin{aligned}
   \delta\psi_M &\approx D_M \epsilon
      -\frac{1}{96}\left(\Gamma_M{}^{P_1P_2P_3}G_{P_1P_2P_3}
         -9\Gamma^{P_1P_2}G_{M P_1P_2}\right)\e^c \\ &\hspace{6cm}
      + \frac{\ii}{192}\Gamma^{P_1P_2P_3P_4}F_{MP_1P_2P_3P_4}\epsilon
      = 0 \\
   \delta\lambda &\approx \ii\Gamma^M P_M \e^c
      + \frac{\ii}{24}\Gamma^{P_1P_2P_3} G_{P_1P_2P_3}\e
      = 0~.
\end{aligned}
\end{equation}
We are working in the formalism where $SU(1,1)$ is realised
linearly. In particular there is a local $U(1)$ invariance and
$Q_M$ acts as the corresponding gauge field. Note that $Q_M$ is a
composite gauge-field with field strength given by \be\label{defQ}
 \dd Q &= -\ii P\wedge P^*~.
\ee Also note that $D$ is the covariant derivative with respect to
local Lorentz transformations and local $U(1)$ transformations.
The spinor $\e$ has $U(1)$ charge 1/2 so that \be
D_M\e=\left(\nabla_M-\frac{\ii}{2}Q_M\right)\e ~.\ee The field $P$
has charge 2, while $G$ has charge 1. We also have the chirality
conditions
$\Gamma_{11}\psi=-\psi$, $\Gamma_{11}\lambda=\lambda$ and
$\Gamma_{11}\e=-\e$.

The equations of motion are\footnote{The sign in the third
equation differs from that of \cite{Schwarz:qr}: we fixed it here
by studying the integrability conditions for supersymmetry, as
discussed in appendix \ref{integrab}.}
\begin{equation}
\label{10d-eom}
\begin{aligned}
   R_{MN} &= P_M P_N^*+P_NP^*_M
      + \frac{1}{96}F_{MP_1P_2P_3P_4}F_N^{~P_1P_2P_3P_4} \\ & \quad
      + \frac{1}{8}\left(
         G_M{}^{P_1P_2}G^*_{NP_1P_2} + G_N{}^{P_1P_2}G^*_{MP_1P_2}
         - \frac{1}{6}g_{MN}G^{P_1P_2P_3}G^*_{P_1P_2P_3} \right)\\
   D^P G_{MNP} &= P^P G^*_{MNP}
      -\frac{\ii}{6}F_{MNP_1P_2P_3}G^{P_1P_2P_3} \\
   D^M P_M &= -\frac{1}{24}G_{P_1P_2P_3}G^{P_1P_2P_3} \\
   F&= *_{10}F~.
\end{aligned}
\end{equation}
We also need to impose the Bianchi identities
\begin{equation}
\begin{aligned}
   DP &= 0 \\
   DG &= -P\wedge G^* \\
   \dd F &= \frac{\ii}{2}G\wedge G^*~.
\end{aligned}
\end{equation}

Note that in the usual string theory variables we have, following \cite{hassan},
\bea
\label{psandqs}
P&=&\frac{\ii}{2}\me^\phi \diff C^{(0)}+\frac{1}{2}
\diff\phi\nn Q&=&-\frac{1}{2} \me^\phi \diff C^{(0)} \eea
and we observe that
the Binachi identity $DP=0$ is identically satisfied. In addition
\be G=\ii\me^{\phi/2}(\tau \diff B-\diff C^{(2)}) \ee (taking into account a
sign difference between our $G$ and that in \cite{hassan}). In
these conventions, according to \cite{hassan}, the $SL(2,\bR)$
action is \be
\tau\to \frac{p\tau+q}{r\tau+s},\qquad \left(\ba{c} C^{(2)} \\
B\ea\right)\rightarrow \left(\ba{cc} p & q \\ r & s \ea\right)
\left(\ba{c} C^{(2)} \\ B\ea\right)\, \ee where $\tau\equiv
C^{(0)}+\ii \me^{-\phi}$, with the Einstein metric and the
five-form left unchanged.


\section{The conditions for supersymmetry in $d=5$}


We consider the most general class of bosonic supersymmetric
solutions of type IIB supergravity with $SO(4,2)$ symmetry. The
$d=10$ metric in Einstein frame is taken to be a warped product
\begin{equation}\label{metansatz}
   \diff s^2_{10} = \me^{2\Delta}\left[\diff s^2(AdS_5)+\diff s^2_5\right]
\end{equation}
where $\diff s^2(AdS_5)$ denotes the metric on $AdS_5$, normalised
so that its Ricci tensor is $-4m^2$ times the metric, and $\diff
s^2_5$ denotes an arbitrary five-dimensional metric on the
internal space $M_5$. $\Delta$ is a real function on this space,
$\Delta\in \Omega^0(M_5,\bR)$. We also take $P\in
\Omega^1(M_5,\bC)$, $Q\in \Omega^1(M_5,\bR)$, $G \in
\Omega^3(M_5,\bC)$ and \be
F=(\vol_{AdS_5}+\vol_5)f\label{genansatz} \ee
where $\vol_5$ denotes the volume form on $M_5$ and $f$ is a real
constant to ensure that the five-form Bianchi identity (or
equation of motion), $\diff F=0$, is satisfied.

For the geometry to preserve supersymmetry it must admit solutions
to the Killing spinor equations \p{10d-susy}. To proceed we
construct the most general ansatz for the spinor $\epsilon$
consistent with minimal supersymmetry in $AdS_5$. As explained in
detail in appendix~\ref{spinors}, $\epsilon$ is constructed from
two spinors, $\xi_i$, of $Spin(5)$ combined with a $Spin(4,1)$
spinor $\psi$ satisfying
$\nabla_{\mu}\psi=\frac{1}{2}m\rho_{\mu}\psi$ on $AdS_5$, where
$\rho_\mu$ generate $\Cliff(4,1)$. After substituting this spinor
ansatz into~\p{10d-susy}, one eventually obtains two differential
conditions
\begin{align}
   D_m \xi_1
      + \frac{\ii}{4} \left(\me^{-4\Delta}f-2m\right)\gamma_m \xi_1
      + \frac{1}{8} \me^{-2\Delta} G_{mnp}\gamma^{np}\xi_2
      &= 0 \label{sone}\\
   \bar{D}_m \xi_2
      -  \frac{\ii}{4} \left(\me^{-4\Delta}f+2m\right)\gamma_m \xi_2
      + \frac{1}{8} \me^{-2\Delta} G_{mnp}^*\gamma^{np}\xi_1
      &= 0 \label{stwo}
\end{align}
and four algebraic conditions
\begin{align}
   \gamma^m\de_m\Delta\xi_1
      - \frac{1}{48}\me^{-2\Delta}\gamma^{mnp}G_{mnp}\xi_2
      - \frac{\ii}{4}\left(\me^{-4\Delta}f-4m\right) \xi_1
      &= 0  \label{sthree}\\
   \gamma^m\de_m\Delta\xi_2
      - \frac{1}{48}\me^{-2\Delta}\gamma^{mnp}G_{mnp}^*\xi_1
      + \frac{\ii}{4}\left(\me^{-4\Delta}f+4m\right)\xi_2
      &= 0 \label{sfour}\\
   \gamma^m P_m \xi_2
      + \frac{1}{24} \me^{-2\Delta} \gamma^{mnp} G_{mnp} \xi_1
      &= 0 \label{sfive}\\
   \gamma^m P_m^* \xi_1
      + \frac{1}{24} \me^{-2\Delta} \gamma^{mnp} G_{mnp}^* \xi_2
      &= 0 \label{ssix}
\end{align}
where $\gamma^m$ generate $\Cliff(5)$ with $\gamma_{12345}=+1$.

It is interesting to consider first the special case where one of
the two spinors, $\xi_2$ say, is identically zero. It is then easy
to see that the warp factor must be constant and related to $f$ by
$f  =  4m \me^{4\Delta}$.  Hence the metrics are direct products
of $AdS_5$ with a five-manifold. In addition we deduce that
\bea\label{steps} G^*_{mnp}\gamma^{np} \xi_1 &=&
G_{mnp}\gamma^{mnp} \xi_1  = 0\nn \gamma^m P_m^* \xi_1&=&0\nn D_m
\xi_1 + \ii \frac{m}{2} \gamma_m \xi_1 & = & 0 ~.\ee The first two
conditions imply\footnote{It is the same calculation that is used
to derive \p{v3} below.} that $G=0$. Next, the third condition
implies $P^2=0$. Writing this out in terms of the axion and
dilaton, using \p{psandqs}, the imaginary part says that $\partial
C^{(0)}\cdot\partial\phi=0$. The equation of motion for $C^{(0)}$
then says that it is harmonic. Now on a compact manifold, which is
the case of most interest for AdS/CFT applications, we deduce that
$C^0$ is constant. The equation of motion for the dilaton then
implies that the dilaton is also constant, for the same reason.
The last condition in \p{steps} then leads us back to the well
known $AdS_5\times X_5$ solutions where $X_5$ is Sasaki--Einstein.

More generally, we can enquire whether it is possible to have
solutions preserving supersymmetry with both $\xi_i$ non-vanishing
but linearly dependent. In fact this is not possible as we show in
appendix \ref{absol}. Note that this implies that the only
solutions with compact $M_5$ having a local $SU(2)$ structure
(determined by supersymmetry), rather than an identity structure
to be considered next, are Sasaki--Einstein.

\subsection{The identity structure}

We now turn to the main focus of the paper: supersymmetric
solutions with $\xi_i$ generically linearly independent. We first note
that, in neighbourhoods where $\xi_i$ are generic, they define,
locally, an identity structure\footnote{An alternative, but
equivalent,  point of view is that the spinors $\xi_1 \otimes \xi_2$ define an
$SU(2)\times SU(2)$ structure on $TM_5\oplus T^*M_5$, 
in the sense of Hitchin \cite{Hitchin:2004ut}
(see also \cite{Jeschek:2004wy,Grana:2005ny}). However, we will not 
adopt this language in the present paper.}, or equivalently, a canonical
orthonormal frame $e^a$. One way to see this is that the set of
spinors $\{\xi_1,\xi_2,\xi_1^c,\xi_2^c\}$ generically form a
complete basis for the spinor representation of $\Spin(5)$.

Equivalently, this structure can easily be seen by noting that
there are six real vectors that can be constructed from two
non-vanishing spinors. These can be written as
\begin{equation}
\label{bilins}
\begin{aligned}
   K^m &\equiv \bar{\xi}_1^c\gamma^m\xi_2 \\
   K^m_3 &\equiv \bar{\xi_2}\gamma^m\xi_1 \\
   K^m_4 &\equiv \tfrac{1}{2}\left(
      \bar{\xi}_1\gamma^m\xi_1 - \bar{\xi}_2\gamma^m\xi_2\right) \\
   K^m_5 &\equiv \tfrac{1}{2}\left(
      \bar{\xi}_1\gamma^m\xi_1 + \bar{\xi}_2\gamma^m\xi_2\right)
\end{aligned}
\end{equation}
where the first two are complex and the last two are real. Since
we are in a five-dimensional space they cannot be linearly
independent. Using Fierz identities one finds that there is a
single linear relation
\begin{equation}
\label{dancheck}
   \epsilon^{ik}\epsilon^{jl}\left(\bar{\xi}_i\xi_j\right)
      \left(\bar{\xi}_k\gamma^m\xi_l\right)
      - 2\re\left(\bar{\xi}_2\xi^c_1\right)
         \left(\bar{\xi}^c_1\gamma^m\xi_2\right)
      = 0
\end{equation}
leaving five independent vectors. From these, given the norms of
the vectors, one can build an orthonormal basis $e^a$ defining the
identity structure. The relation between the vectors and a
particular useful basis $e^a$ is given in appendix~\ref{basis}.
Again by Fierz identities, one can write the norms of the vectors
in terms of the six independent scalar bilinears. These can be
parameterised as
\begin{equation}
\label{scalars}
\begin{aligned}
   A &\equiv \tfrac{1}{2}\left(
      \bar{\xi}_1\xi_1 + \bar{\xi}_2\xi_2\right) \\
   A\sin\zeta &\equiv \tfrac{1}{2}\left(
      \bar{\xi}_1\xi_1 - \bar{\xi}_2\xi_2\right) \\
   S &\equiv \bar{\xi}^c_2\xi_1 \\
   Z &\equiv \bar{\xi}_2\xi_1 \\
\end{aligned}
\end{equation}
where the first two are real and the second two are complex. In
summary, these vector and scalar bilinears define the identity
structure.

We now aim to find the conditions on the identity structure and on
the fluxes that are equivalent to supersymmetry. This calculation
falls into two parts. First one considers the differential
conditions~\p{sone} and~\p{stwo}. This is equivalent to giving the
intrinsic torsion, or here since we have an identity structure,
the torsion itself, in terms of the flux, $f$, $m$ and the warp factor
$\Delta$. The  same
information is contained in the exterior derivatives of the
canonical orthonormal frame $e^m$, which is in turn encoded in the
exterior derivatives of the vector and the scalar bilinears. The
second step is then to find necessary and sufficient constraints
on the structure due to the algebraic
conditions~\p{sthree}--\p{ssix}.

\subsection{Torsion conditions}
\label{sec:torsion}

In calculating the torsion conditions it is convenient to work not
with the exterior derivatives of a particular orthonormal basis
$e^m$, but rather the exterior derivatives of the vector and
scalar bilinears defined above, which are completely equivalent.
The results of appendix~\ref{basis} then provide a translation to
$e^m$ if required.

We start by calculating the derivatives of the scalar bilinears.
Making use of the algebraic conditions~\p{sthree}--\p{ssix}, one
finds first that $A$ is constant, $\dd A = 0$. Thus we can
consistently set
\begin{equation}
\label{eq:dA}
      A = 1~.
\end{equation}
The remaining scalars then satisfy
\begin{align}
   \dd(\me^{4\Delta} \sin\zeta) &= 0 \label{s1} \\
   \me^{-4\Delta}\diff (\me^{4\Delta} S) &= 3\ii m K \label{s2} \\
   \me^{-2\Delta}D (\me^{2\Delta}Z) &= -PZ^* \label{s3}~.
\end{align}

Next we turn to the vectors. Again with some judicious use of the
conditions~\p{sthree}--\p{ssix}, after some work we find that the
differential constraints~\eqref{sone} and~\eqref{stwo} imply
\begin{align}
   \diff (\me^{4\Delta}K) &= 0 \label{vK} \\
   \me^{-6\Delta} D(\me^{6\Delta} K_3)
      &= P\wedge K_3^* - 4\ii m W - \me^{-2\Delta}*G \label{v3}\\
   \me^{-4\Delta} \diff(\me^{4\Delta} K_4)
      &= -2mV \label{v4}\\
   \me^{-8\Delta}\diff (\me^{8\Delta} K_5)
      &= \me^{-4\Delta}fV - 6 m U - \re\left(\me^{-2\Delta}Z^**G\right)
      \label{v5}
\end{align}
where we have introduced the two-forms
\begin{equation}
\label{eq:2forms}
\begin{aligned}
   \ii U_{mn} &\equiv \tfrac{1}{2}\left(
      \bar{\xi}_1\gamma_{mn}\xi_1 + \bar{\xi}_2\gamma_{mn}\xi_2
      \right) \\
   \ii V_{mn} &\equiv \tfrac{1}{2}\left(
      \bar{\xi}_1\gamma_{mn}\xi_1 - \bar{\xi}_2\gamma_{mn}\xi_2
      \right) \\
   W_{mn} &\equiv -\bar{\xi}_2\gamma_{mn}\xi_1
\end{aligned}
\end{equation}
which can, of course, be rewritten in terms of the basis $e^m$
(see appendix~\ref{basis}) and hence the scalar and vector
bilinears. Doing so, or using Fierz identities, and given that $A=1$,
one finds the identity
\begin{equation}
\label{UVrel}
   \sin\zeta  V  - U  - \tfrac{\ii}{2} K^*\wedge  K + \re [\ii Z^* W]
      = 0 \ .
\end{equation}

We note first that the first differential condition~\eqref{vK} is
in fact implied by the scalar condition~\eqref{s2}. Next we recall
that the six vector bilinears are not independent. The linear
relation~\eqref{dancheck} implies that
\begin{equation}
\label{linrel}
   K_5 = \sin\zeta K_4 + \re [Z^* K_3] - \re [S^*K]
\end{equation}
where we have again used the fact that $A=1$. Taking the exterior
derivative of~\p{linrel} and comparing with $\diff K_5$ in~\p{v5}
gives the consistency condition
\begin{equation}
\label{check}
   (\me^{-4\Delta}f+2m\sin\zeta) V
      =  6m U - 4m \re [\ii Z^* W] +3\ii m K^*\wedge  K~.
\end{equation}
However, the two-forms above are linearly related; in particular
they obey the identity~\eqref{UVrel}. To be consistent with this
identity we require, first, that
\begin{equation}
\label{fxf}
   \me^{-4\Delta}f = 4m \sin\zeta
\end{equation}
fixing the integration constant in the differential
condition~\eqref{s1} (In fact, it is straightforward to show this
relation holds, directly from the algebraic
constraints~\eqref{sthree} and~\eqref{sfour}). Secondly, we also
require $\re [\ii Z^* W]= 0$. Using the explicit expression for $W$
(see appendix~\ref{basis}), it is easy to show that this implies
the important condition
\begin{equation}
   Z = 0~.
\end{equation}
This condition simplifies considerably the algebraic and
differential conditions obeyed by the bilinears.

In summary, the torsion conditions~\eqref{sone}
and~\eqref{stwo} are equivalent\footnote{Note we have also
used some of the algebraic conditions~\eqref{sthree}--\eqref{ssix}.}  to
\begin{equation}\label{ayzed}
   \me^{-4\Delta}f = 4m \sin\zeta, \qquad A = 1, \qquad Z = 0
\end{equation}
together with the differential conditions
\begin{align}
   \me^{-4\Delta}\diff (\me^{4\Delta} S) &= 3\ii m K \label{S-cond} \\
   \me^{-6\Delta} D(\me^{6\Delta} K_3)
      &= P\wedge K_3^* - 4\ii m W - \me^{-2\Delta}*G \label{K3-cond}\\
   \me^{-8\Delta}\diff (\me^{8\Delta} K_5)
      &= 4m\sin\zeta V - 6 m U \label{K5-cond}~.
\end{align}
(We drop the $\dd K_4$ condition since the linear dependence means
it is implied by the other vector bilinear conditions.) As
expected, starting with the work \cite{Gauntlett:2001ur} and
others following this (in particular, see \cite{Gauntlett:2002fz,Gauntlett:2003cy,Martelli:2003ki,
Hackett-Jones:2004yi,Cascales:2004qp}), we note that these differential conditions
are written in a form reminiscent of  ``generalized
calibrations''~\cite{Gutowski:1999iu,Gutowski:1999tu}.

\subsection{Algebraic conditions}
\label{sec:alg-conds}

Next we turn to the algebraic
conditions~\eqref{sthree}--\eqref{ssix}. We would like to find the
equivalent algebraic conditions relating the identity structure,
$P$, $G$, $f$, $m$, and $\Delta$. The simplest way to do this is to note
that, as mentioned above, generically the set
$\eta_\alpha\in\{\xi_1,\xi_2,\xi_1^c,\xi_2^c\}$ form a complete
basis in the $\Spin(5)$ spinor representation space. Thus we can
construct the identity operator
\begin{equation}
\label{complete}
   \id = \eta_\alpha (m^{-1})^{\alpha\beta} \bar{\eta}_\beta
\end{equation}
where $m_{\alpha\beta}=\bar{\eta}_\alpha\eta_\beta$.

Next, using the fact that
$\gamma_{mnp}G^{mnp}=-3*G_{mn}\gamma^{mn}$ one rewrites the
algebraic conditions in the form
\begin{equation}
   \me^{-2\Delta}{*G}_{mn}\gamma^{mn}\eta_\alpha
      = \Sigma_\alpha{}^\beta\eta_\beta \ .
\end{equation}
Using the completeness relation~\eqref{complete}, we see that the
algebraic conditions are equivalent to an operator equation
\begin{equation}
   \me^{-2\Delta}*G_{mn}\gamma^{mn}
      = \Sigma_{\alpha}{}^\gamma (m^{-1})^{\alpha\beta}
        \eta_\gamma\bar{\eta}_\beta~.
\end{equation}
Performing Fierz identities on $\eta_\gamma\bar{\eta}_\beta$ one
gets three types of relations. From the $\id$ coefficient one
finds
\begin{equation}
   \label{Pharm}
   i_{K_3^*} P = 2 i_{K_3} \dd\Delta \ .
\end{equation}
The $\gamma^m$ coefficient gives three additional conditions
\begin{gather}
   i_{K_5} \dd\Delta = 0    \label{Delta-killing}\\
   i_{K_5} P = 0    \label{P-killing} \\
   \me^{-4\Delta}f = 4m \sin\zeta   \label{fixf}
\end{gather}
where the final expression has already appeared as a consistency
condition~\eqref{fxf}. The $\gamma^{mn}$ coefficient meanwhile
gives an expression for the flux $*G$ 
\begin{equation}
\label{flux}
\begin{aligned}
   (\cos^2\zeta-|S|^2) &\, \me^{-2\Delta} *G
      \\ &=
      2P \wedge K_3^*
      - \left(4\dd\Delta + 4\ii m K_4
         - 4 \ii m\sin\zeta K_5\right) \wedge K_3
      \\ & \quad
      + 2 * \left( P\wedge K_3^*\wedge K_5
         - 2\dd\Delta\wedge K_3\wedge K_5 \right)~.
\end{aligned}
\end{equation}
In deriving this last expression one uses the identities
\begin{equation}
\begin{aligned}
   S^*\xi_1^c\gamma_{(2)}\xi_1
      &= \left(1+\sin\zeta\right) W - (K_4+K_5)\wedge K_3 \\
   S^*\xi_2^c\gamma_{(2)}\xi_2
      &= \left(1-\sin\zeta\right) W^* - (K_4-K_5)\wedge K_3^*~.
\end{aligned}
\end{equation}
For completeness, we also note that, with $Z=0$, the two-forms
$U$, $V$ and $W$ are given by 
\begin{equation}
\label{UVWexp}
\begin{aligned}
   U &= \frac{1}{2(\cos^2\zeta-|S|^2)}\big(
       \ii \sin\zeta K_3 \wedge K_3^*
       + \ii K \wedge K^* - 2\im S^*K\wedge K_5\big) ,\\
   V &= \frac{1}{2\sin\zeta(\cos^2\zeta-|S|^2)}\big(
       \ii \sin\zeta K_3\wedge K_3^*
       \\ & \qquad\qquad\qquad  {}
       + \ii [\sin^2\zeta+|S|^2] K\wedge K^*
       - 2\im S^*K\wedge K_5\big) ,\\
   W & = \frac{1}{\sin\zeta(\cos^2\zeta-|S|^2)}\big(
       \cos^2\zeta K_5
       + \re S^*K
       + \ii\sin\zeta\im S^*K\big)\wedge K_3 .
\end{aligned}
\end{equation}
%

In summary, the algebraic conditions~\eqref{Pharm}--\eqref{flux},
together with~\p{ayzed} and~\eqref{S-cond}--\eqref{K5-cond}, are
equivalent to the Killing spinor
equations~\eqref{sone}--\eqref{ssix}.

\subsection{The Killing vector $K_5$}

We now show that $K_5$ is a Killing vector and, moreover, it
generates a symmetry of the full solution. This corresponds to the
fact that the dual $D=4$ superconformal field theories have a
global $U(1)_R$ symmetry. While the Killing condition is implied
by the necessary and sufficient
conditions~\eqref{S-cond}--\eqref{K5-cond}
and~\eqref{Pharm}--\eqref{flux}, the simplest derivation is
directly from the spinor conditions~\p{sone}--\p{ssix}.
Calculating $\nabla K_5$, one can easily show that the symmetric part
vanishes and hence $K_{5}$ is Killing.

We next compute its action on the remaining bosonic fields.
From~\eqref{Delta-killing} we immediately see that
\begin{equation}
   \mathcal{L}_{K_5}\Delta = 0
\end{equation}
and hence, by~\eqref{fixf}, $\mathcal{L}_{K_5}\zeta=0$.
From~\eqref{P-killing}, given the expression~\p{psandqs} for $P$,
one also immediately has
\begin{equation}
   \mathcal{L}_{K_5}\phi = \mathcal{L}_{K_5}C^{(0)} =0
      \ \Leftrightarrow\ \mathcal{L}_{K_5}P = 0~.
\end{equation}
Finally, we need to consider $\mathcal{L}_{K_5}G$. This can be
calculated directly from the expression~\eqref{flux} for $*G$. To
do so we need to know the action of the Lie derivative
$\mathcal{L}_{K_5}$ on the scalar and vector bilinears. One finds
that all the bilinears are invariant except for
\begin{equation}
\label{lies}
   \mathcal{L}_{K_5} S = -3 \ii mS ~,
\end{equation}
(and hence also $\mathcal{L}_{K_5}K=-3\ii mK$). This implies
$\mathcal{L}_{K_5}(SS^*)=0$ and thus, from~\eqref{flux},
\begin{equation}
   \mathcal{L}_{K_5}G = 0~.
\end{equation}
We thus see that the Killing vector $K_5$ does indeed generate a
symmetry of the full solution.


\subsection{Equations of motion}

We now show that the conditions we have derived for supersymmetry
automatically imply the equations of motion and the Bianchi
identities.

We first recall that $DP=0$ follows automatically from the
expression for $P$ in terms of the variables \p{psandqs}. Also,
our ansatz has $\dd F=0$ by construction. Next,
from~\eqref{K3-cond} (and using~\p{fixf}) we find
\begin{equation}
   D(\me^{4\Delta}*G) = \me^{4\Delta}P\wedge * G^* - \ii f  G
\end{equation}
which is just the $G$ equation of motion. The easiest way to show
that the $G$ Bianchi identity is also satisfied is to derive a
differential condition for $W$ directly from spinor
conditions~\eqref{sone}--\eqref{ssix}. One finds
\begin{equation}
   D (\me^{6\Delta}W)
      = -\me^{6\Delta}P\wedge W^* + (f/4m) G ~.
\end{equation}
Taking a  derivative with $D$ then reproduces the Bianchi
identity for $G$.

In appendix~\ref{integrab} we consider the integrability
conditions for the Killing spinor equations. Assuming that the
$P$, $G$ and $F$ Bianchi identities, together with the $G$ equation
of motion are satisfied, one finds that a supersymmetric
background necessarily satisfies the $P$ equations of motion.
Moreover, all but one component of Einstein's equations is
automatically satisfied. In  appendix~\ref{integrab} we also show
that this component is satisfied in the present case, so we can
conclude that:

\begin{quote}
   \textsl{For the class of solutions with metric of
   the form \p{metansatz} and fluxes respecting $SO(4,2)$ symmetry,
   all the equations of motion and Bianchi identities are implied by
   supersymmetry.}
\end{quote}
A similar situation was found to hold for the supersymmetric
$AdS_5$ M-theory solutions of~\cite{gmsw}. Clearly, this is very
useful for constructing solutions. In fact, it is often the
Bianchi identities that are the difficult equations to satisfy.

\subsection{Summary}
Let us end by summarising the necessary and sufficient
conditions for the generic supersymmetric solution with metric of
the form \p{metansatz} and fluxes respecting $SO(4,2)$ symmetry.
$M_5$ must admit an identity structure defined by two spinors
$\xi_1, \xi_2$ which determine the preserved supersymmetry. The
scalars $A$ and $Z$ defined in \p{scalars} are given by $A=1$ and
$Z=0$. The identity structure can then be specified by a real
vector $K_5$, and two complex vectors $K, K_3$ defined in
\p{bilins}, along with a real scalar $\zeta$ and a complex scalar
$S$ defined in \p{scalars}. These satisfy the following conditions
\begin{align}
   \me^{-4\Delta}\diff (\me^{4\Delta} S) &= 3\ii m K \label{S-summ} \\
   \me^{-6\Delta} D(\me^{6\Delta} K_3)
      &= P\wedge K_3^* - 4\ii m W - \me^{-2\Delta}*G \label{K3-summ}\\
   \me^{-8\Delta}\diff (\me^{8\Delta} K_5)
      &= 4m\sin\zeta V - 6 m U \label{K5-summ}
\end{align}
and the additional algebraic constraint
\begin{equation}
\label{holo-sum}
   i_{K_3^*} P = 2\,i_{K_3} \dd\Delta \ .
\end{equation}
The five-form flux is given by \p{genansatz} with
\begin{equation}
\label{f-summ}
   f = 4m \me^{4\Delta}\sin\zeta
\end{equation}
while the three-form flux is given by 
\begin{equation}
\label{flux-summ}
\begin{aligned}
   \left(\cos^2\zeta-|S|^2\right) &\, \me^{-2\Delta} *G
      \\ &
      = 2P \wedge K_3^*
      - \left(4\dd\Delta + 4\ii m K_4
         - 4 \ii m\sin\zeta K_5\right) \wedge K_3
      \\ & 
      + 2 * \left( P\wedge K_3^*\wedge K_5
         - 2\dd\Delta\wedge K_3\wedge K_5 \right)~.
\end{aligned}
\end{equation}
 The metric
can be written (using results in appendix~\ref{basis})

\begin{equation}
\begin{aligned}
\diff s^2 & = \frac{(K_5)^{2}}{\sin^2\zeta+|S|^2}
    + \frac{K_3\otimes K_3^*}{\cos^2\zeta-|S|^2}
    + \frac{|S|^2}{\cos^2\zeta-|S|^2}
        \left(\im{S^{-1}K}\right)^{2}
        \\ & \qquad
    + \frac{|S|^2}{\sin^2\zeta}\;
        \frac{\sin^2\zeta+|S|^2}{\cos^2\zeta-|S|^2}
        \left(\re{S^{-1}K}
            + \frac{1}{\sin^2\zeta+|S|^2}K_5\right)^{2}~.
\end{aligned}
\end{equation}

The conditions imply that $K_5$ is a Killing vector field that
generates a symmetry of the full solution:
$\mathcal{L}_{K_5}\Delta=i_{K_5}P=\mathcal{L}_{K_5}G=0$.
Furthermore, all equations of motion and the Bianchi identities
are satisfied.


\section{Reducing the conditions}


It is now useful to introduce some convenient local coordinates and
hence reduce the conditions to a simpler set. We will first reduce on
the Killing direction $K_5$ and then use the condition~\eqref{S-summ}
to write the resulting four-dimensional metric as a product of a
one-dimensional metric and a three-dimensional metric
$\tilde{g}$. The problem then reduces to a set of conditions on
the local identity structure on $\tilde{g}$. 

We begin by choosing a coordinate $\psi$ that is adapted to the
Killing direction $K_5$. As a vector, we write 
\be\label{firstexpkil}
 K_5^{\#}  =  3m\frac{\partial}{\partial \psi}
\ee
and therefore as a one-form
\be\label{lowkil}
 K_5 =  \frac{1}{3m} \cos^2\eta\;(\dd\psi + \rho)~,
\ee
where $\cos\eta$ is the norm of $K_5$, given by
$\cos^2\eta=\sin^2\zeta+ |S|^2$. (Note that in the conventions of
appendix~\ref{basis} $\eta=2\phi$.) 
%
%

Let us now turn to~\eqref{S-summ}. The dependence of $S$ on $\psi$ is
given by the Lie derivative~\eqref{lies} which is solved by
$S=\me^{-\ii\psi}\hat{S}$, where the complex scalar $\hat{S}$ is
independent of $\psi$. Noting that there is a gauge freedom in
shifting the coordinate $\psi$ by a function of the remaining
coordinates, we take $\hat S$ to be real, without loss of
generality. It is then natural to introduce a coordinate $\lambda$
such that 
\begin{equation}\label{esscond}
   S = \sin\zeta\,\lambda\,\me^{-\ii\psi}~,
\end{equation}
where the factor of $\sin\zeta$ is added for convenience so
that~\eqref{S-summ} now reads, given~\eqref{f-summ},
\begin{equation}
   S^{-1}K = -\frac{1}{3m}\left(\dd\psi+\ii\dd\ln\lambda\right)~. 
\end{equation}
Note that in these coordinates we have
\be
\label{zeta-redef} 
   \sin\zeta = \frac{\cos\eta}{(1+\lambda^2)^{1/2}}~, 
\ee
and it is convenient to switch to $\eta$, $\psi$  and $\lambda$
instead of the scalars $\zeta$ and $S$. 

For convenience let us also define a new complex one-form $\sigma$ by 
\begin{equation}
   K_3=\sigma/3m~.
\end{equation}
Using the results contained in appendix~\ref{basis}, one can write the
underlying orthonormal frame as 
\begin{equation}
\begin{aligned}
   3m\, e^1 &= \cos\eta\,(\diff \psi+\rho)~, \\
   3m\, e^2 &= \lambda\cot\eta\;\rho~, \\ 
   3m(-\ii e^3 + e^4) &= \frac{1}{\sin\eta}\; \sigma~, \\
   3m\, e^5 &= \frac{\cot\eta}{(1+\lambda^2)^{1/2}}\; \diff\lambda~.
\end{aligned}
\end{equation}
Now since $\mathcal{L}_{K_5}\rho=\mathcal{L}_{K_5}\sigma$ we can
always choose coordinates $(\psi,\lambda,x^i)$ such that $\rho$ and
$\sigma$ are independent of $\dd\lambda$. (One first reduces on the
Killing direction to a four-dimensional metric, independent of $\psi$,
spanned by $e^2$, $e^3$, $e^4$ and $e^5$. Then given
$e^5\sim\dd\lambda$ one can always make a four-dimensional coordinate
transformation such that there are no cross-terms $\dd\lambda\dd x^i$
in the metric.) Thus the five-dimensional metric has the form
\begin{equation}
\label{firstexpmet}
   9m^2\,\diff s^2_5 = \cos^2\eta\, (\diff \psi+\rho)^2 
      + \frac{\cot^2\eta}{1+\lambda^2}\,\diff \lambda^2 
      + \tilde{g}_{ij}(\lambda,x^i)\,\dd x^i\dd x^j~,
\end{equation}
where the three-dimensional metric $\tilde{g}$ is given in
terms of $\sigma$ and $\rho$
\begin{equation}
   \tilde{g}_{ij}(\lambda,x^i)\,\dd x^i\dd x^j
      = \lambda^2\cot^2\eta\, \rho^2 
       + \frac{\sigma\otimes\sigma^*}{\sin^2\eta}~.
\end{equation}

In summary, we have reduced the problem to a three-dimensional metric
$\tilde{g}$ with a local identity structure given by
$(\rho,\sigma,\sigma^*)$ which also depends on the coordinate $\lambda$. In
addition, there is one remaining scalar $\eta$. (Note that $\Delta$ is
given in terms of $\eta$ and $\lambda$ using~\eqref{f-summ}
and~\eqref{zeta-redef}.). In making this reduction we have
used~\eqref{S-summ} and the fact that $K_5$ is Killing. It remains to
translate the remaining conditions~\eqref{K3-summ} and~\eqref{K5-summ}
into conditions on $\rho$ and $\sigma$. 

Let us first split 
\begin{equation}
   \diff = \tilde{\diff} + \dd\lambda\frac{\de}{\de \lambda}
      + \dd\psi\frac{\de}{\de\psi}
\end{equation}
where $\tilde{\dd}=\dd x^i\de/\de x^i$. Similarly we write 
\begin{equation}
   P = \tilde{P} + P_\lambda\dd\lambda~,
\end{equation}
recalling that $i_{K_5}P=0$. Writing $\de_\lambda=\de/\de\lambda$, the
condition~\eqref{K5-summ} is equivalent to 
\begin{equation}
\label{rho-eqs}
\begin{aligned}
   \de_\lambda \rho &= 
      -\frac{2(1+2\sin^2\eta)\lambda}{3\sin^2\eta(1+\lambda^2)}\,\rho~,
      \\
   \tilde\dd\rho &=
      -\frac{\ii}{3\sin^2\eta\cos\eta(1+\lambda^2)^{1/2}}
         \,\sigma\wedge\sigma^*~.
\end{aligned}
\end{equation}
Similarly the condition~\eqref{K3-summ} reduces to
\begin{equation}
\label{sigma-eqs}
\begin{aligned}
   \sin^2\eta\, \me^{-6\Delta}D_\lambda(\me^{6\Delta}\sigma)
      &= \left(4\de_\lambda\Delta
         - \frac{4\lambda\cos^2\eta}{3(1+\lambda^2)}\right)\sigma
        - \left(1+\cos^2\eta\right)P_\lambda\sigma^* \\ &\qquad
        - \frac{2\cos^2\eta}{\sin\eta(1+\lambda^2)^{1/2}}\,
           \tilde{*}\left( 2\tilde{\dd}\Delta\wedge\sigma
              - \tilde{P}\wedge\sigma^* \right)~, \\
   \sin^2\eta\, \me^{-6\Delta}\tilde{D}(\me^{6\Delta}\sigma)
      &= 4\tilde{\dd}\Delta \wedge \sigma
        - \left(1+\cos^2\eta\right)\tilde{P} \wedge \sigma^* 
        \\ &\qquad
        - 2\ii\lambda(1+\lambda^2)^{1/2}\cos\eta\,
           \rho\wedge\left( 2\de_\lambda\Delta\,\sigma
              + P_\lambda\,\sigma^* \right)~.
\end{aligned}
\end{equation}
The only remaining condition is the algebraic
relation~\eqref{holo-sum} which reads
\begin{equation}
\label{Delta-eq}
   i_{\sigma^*}P = 2 i_\sigma \dd\Delta~. 
\end{equation}
In summary, one needs to solve~\eqref{rho-eqs} and~\eqref{sigma-eqs}
subject to~\eqref{Delta-eq}. This concludes our analysis of the most 
general $AdS_5$ geometries arising in type IIB supergravity.

\section{A simplifying ansatz}


In order to find explicit solutions to these equations we will now
make a particular, very natural, ansatz. First we assume that the
dilaton is constant and the axion zero, $P=0$. Then we assume that the
one-forms $(\rho,\sigma)$ are (locally) proportional to the
left-invariant one-forms on $S^3$, that is  
\begin{equation}
\begin{aligned}
   \rho &= A \sigma_3~, \\
   \sigma &= B (\sigma_2 - \ii\sigma_1)~.
\end{aligned}
\end{equation}
(Note that with this choice $(\sigma_1,\sigma_2,\sigma_3)$ define the
same orientation as $(e^2,e^3,e^4)$.) Explicitly we can introduce
coordinates $\sigma_3= \diff y-\cos\alpha\diff\beta$ and
$\sigma_1=-\sin y \diff\alpha-\cos y\sin\alpha\diff\beta$,
$\sigma_2=\cos y \diff\alpha-\sin y\sin\alpha\diff\beta$. 
In addition we assume that the functions $A$, $B$ and $\eta$ all
depend only on $\lambda$. As we will see this ansatz means the metric
has a local $\SU(2)\times U(1)\times U(1)$ isometry group. 

We find that the entire analysis then boils down to solving a
second-order non-linear ordinary differential equation. Furthermore,
we find one exact solution to this ODE which after a
change of coordinates turns out to be precisely a solution first found
by Pilch and Warner~\cite{pw}. Our numerical investigations of the ODE
lead to a one parameter family of local solutions, which do not extend
to globally defined solutions, as we will discuss. 

We start by introducing two functions
\begin{equation}
\begin{aligned}
   h &= - A (1+\lambda^2)~, \\
   g &= \frac{1}{\sin\zeta} = \frac{(1+\lambda^2)^{1/2}}{\cos\eta}~. 
\end{aligned}
\end{equation}
To satisfy the $\tilde{\dd}$-equation in~\eqref{rho-eqs} one requires
\begin{equation}
   B = \left[ \frac{3h(g^2-1-\lambda^2)}{2g^3}\right]^{1/2}~.
\end{equation}
This implies that the metric takes the form 
\begin{equation}
\label{fourbase}
\begin{aligned}
   9m^2\,\dd s_5^2 &= \frac{1+\lambda^2}{g^2}\left(\dd \psi
      -\frac{h}{1+\lambda^2}\sigma_3\right)^2 \\
      & \qquad + \frac{1}{g^2-1-\lambda^2}\left( 
         \diff \lambda^2 
         + \frac{\lambda^2}{1+\lambda^2}h^2\sigma_3^2 \right)
      + \frac{3h}{2g} (\sigma_1^2+\sigma_2^2)~,
\end{aligned}
\end{equation}
and it is clear that the metric has a local $\SU(2)\times U(1)\times
U(1)$ isometry group. (Note that $\sigma_1^2+\sigma_2^2$ is just the
round metric on $S^2$.) The $\de_\lambda$-equations
in~\eqref{rho-eqs} and the conditions~\eqref{sigma-eqs} all reduce to
a pair of coupled first-order differential equations for $g$ and $h$,
namely
\begin{equation}
\label{odes}
\begin{aligned}
   \dot h &= - \frac{2\lambda h}{3}\frac{1}{g^2-1-\lambda^2}~, \\
   \dot g &= \frac{1}{\lambda h}(g^2-1-\lambda^2)~.
\end{aligned}
\end{equation}
where the dot denotes $\de_\lambda$. These are equivalent to a second
order ODE for $g$, which reads
\begin{equation}
\label{gode}
   \ddot g \lambda (g^2-1-\lambda^2) 
     + \dot g (g^2 -1 +\tfrac{1}{3}\lambda^2
     - 2\lambda g \dot g ) = 0 ~.
\end{equation}
Any solution to these equations gives rise to a (local) supersymmetric
solution with an $AdS_5$ factor and non-trivial 3-form flux. 


For completeness we note that the flux is given by
\begin{equation}
\label{fluxforspans}
\begin{aligned}
   G &= \left(\frac{f}{4m}\right)^{1/2}mg^{1/2}\Bigg[
       \frac{4\lambda^2gh-3(1+\lambda^2)(g^2-1-\lambda^2)}
            {\lambda(1+\lambda^2)^{1/2}gh}\,e^{15}
       \\ & \qquad\qquad\qquad
       -4\frac{(g^2-1-\lambda^2)^{1/2}}{g(1+\lambda^2)^{1/2}}\,e^{25} 
       + 3\ii\frac{g^2-1-\lambda^2}{\lambda h}\,e^{12}
       \Bigg]\wedge(e^4-\ii e^3)~.
\end{aligned}   
\end{equation}
One can also integrate this expression to give the complex potential
$A$ in a relatively simple form~\eqref{Aform}. Note also that the
equations~\eqref{odes} are symmetric under $\lambda\rightarrow
-\lambda$, $g\rightarrow -g$ and $h\rightarrow -h$.

\subsection{Pilch--Warner solution}

We managed to find a single analytic solution to~\eqref{odes} given by 
\bea 
g &= &1+\tfrac{1}{\sqrt{3}}\lambda\nn  
h &=& 2\left(1-\tfrac{1}{\sqrt{3}}\lambda\right)~.
\eea
We now show that this is locally the same solution first found by
Pilch and Warner \cite{pw}. To see this we take the range of $\lambda$
to be $0\le\lambda\le\sqrt 3$ and change coordinates via
\bea 
\lambda&=&\sqrt{3}\sin^2\theta\nn 
\psi & = & 2\phi\nn
y&=&\gamma+2\phi~.
\eea
Also define corresponding set of left-invariant forms $\hat{\sigma}_3=
\diff \gamma-\cos\alpha \diff\beta$ and $\hat{\sigma}_1=-\sin\gamma
\diff\alpha-\cos\gamma\sin\alpha \diff\beta$,
$\hat{\sigma}_2=\cos\gamma \diff\alpha-\sin\gamma\sin\alpha
\diff\beta$.

Then the metric can be written as \bea (9m^2)\dd s^2_5&=&
6\dd \theta^2+\frac{6\sin^2(2\theta)}{(3-\cos(2\theta))^2}\hat{\sigma}_3^2
+\frac{6\cos^2\theta}{3-\cos(2\theta)}(\hat{\sigma}_1^2+
\hat{\sigma}_2^2)\nn &+&4\left[\dd
\phi+\frac{2\cos^2\theta}{3-\cos(2\theta)}\hat{\sigma}_3\right]^2 \eea
which is the form of the metric as written by Pilch and Warner.
Note that in the new coordinates the canonical Killing vector takes the form
$\partial_\psi=(1/2)\partial_\phi-\partial_\gamma$.
The warp factor is given by
\be
\me^{2\Delta}=\left(\frac{f}{4m}\right)^{1/2}\frac{(3-\cos2\theta)^{1/2}}{\sqrt
2}~. \eea
In the new orthonormal frame with $\hat e^1\propto
[\diff\phi+...]$, $\hat e^2\propto \hat\sigma_3$, $\hat
e^3\propto\hat\sigma_1$, $\hat e^4\propto \hat\sigma_2$, the flux
is given by
\bea G=m{\sqrt 3}\me^{2\Delta}\me^{2i\phi}\left(\hat
e^1+\ii\frac{ {\sqrt 2} \sin 2\theta}{{\sqrt 3}(3-\cos2\theta)}\hat
e^5\right)(\hat e^5-\ii\hat e^2)(\hat e^4-\ii\hat e^3) \eea
which coincides with \cite{pw} (up to possible factors).

The fibration defined by $\partial_\phi$ defines, locally, a
four-dimensional base space. However, it is easy to see that there
is no choice of the range of coordinates $\phi$ and $\hat\psi$
that makes it a regular four-dimensional manifold. The same is
true of the base space defined by the foliation using
$\partial_y$. We therefore introduce a new set of coordinates
defined by \bea \phi&=&\delta\nn \gamma&=&\gamma'+\delta~. \eea
Then the metric takes the form 
\bea 
   9m^2 \diff s^2_5&=&
      6\diff\theta^2 
      + \frac{12\sin^22\theta}{35-3\cos^22\theta}(\sigma_3')^2
      + \frac{3(1+\cos2\theta)}{3-\cos2\theta}\left[
         (\sigma_1')^2+(\sigma_2')^2 \right] \nn
      &&\qquad\qquad 
      {}+ \frac{2(35-3\cos^22\theta)}{(3-\cos2\theta)^2}
         (\diff\delta+A)^2~, 
\eea 
where $\sigma_i'$ are the left invariant one-forms $\hat{\sigma}_i$
above with $\gamma$ replaced with $\gamma'$ and the one-form $A$ is
given by 
\be
A=\frac{(1+\cos2\theta)(11-3\cos(2\theta))}{(35-3\cos^2(2\theta))}\sigma_3'
~.\ee If we choose the period of $\gamma'$ to be $4\pi$ then it is
not difficult to see that the four-dimensional
base orthogonal to $\partial_\delta$ is diffeomorphic to
$\mathbb{C}P^2$. In particular at $\theta=0$ the metric has a
two-sphere bolt, with normal neighbourhoood being that of the
chiral spin bundle of $S^2$, while at $\theta=\pi/2$ the metric
has a NUT, {\it i.e.} it smoothly approaches $\mathbb{R}^4$. The
full space is obtained by gluing these together which gives
$\mathbb{C}P^2$. Furthermore, we note that the single non-trivial
two-cycle of the base space is represented by the two-sphere bolt at $\theta=0$.
We next analyse the fibre direction $\partial_\delta$. First note
that the norm of this Killing vector field is nowhere vanishing.
The one-form $A$ is a bona-fide connection one-form; its first
Chern class, defined by the integral of $\diff A/(2\pi)$ over the
two-sphere bolt, is one. After recalling the Hopf fibration of
$S^5$ over $\mathbb{C}P^2$, we conclude that if we choose the
period\footnote{For completeness we note that the periodicities of
$\delta$ and $\gamma'$ imply that $y$ is a periodic coordinate
with period $4\pi$ while the range of $\psi$ is $4\pi$.} of
$\delta$ to be $2\pi$ the topology of the five-dimensional space
is in fact $S^5$.

\subsection{Numerical analysis}

A numerical investigation of the ODE seems to reveal a continuous
family of solutions containing the PW solution and all with
topology $S^5$. We summarise the main points first and then
discuss  how the three-form flux and the spinors are not
globally defined.

Following on from our discussion of the PW solution, we first consider the general coordinate transformation
\bea
\psi&=&2\delta\nn y&=&\gamma'+c\delta ~.\ee
The metric then takes the form
\be (9m^2) \dd s^2= A[\dd\delta+D\sigma_3']^2
+\frac{\dd\lambda^2}{g^2-1-\lambda^2}
+Q(\sigma_3')^2+\frac{3h}{2g}((\sigma_1')^2+(\sigma_2')^2) \ee
where
\bea
A&=&\frac{-4\lambda^4-8\lambda^2+4\lambda^2g^2+4\lambda^2hc+4g^2-4+4hc+h^2c^2g^2-4hc g^2-h^2c^2}
{(g^2-1-\lambda^2)g^2}
\nonumber\\[2mm]
Q&=&\frac{4\lambda^2h^2}
{-4\lambda^4-8\lambda^2+4\lambda^2g^2+4\lambda^2hc+4g^2-4+4hc+h^2c^2g^2-4hc g^2-h^2c^2}
\nonumber\\[3mm]
D&=&\frac{(2\lambda^2-2g^2+2-hc+hc g^2)h}
{-4\lambda^4-8\lambda^2+4\lambda^2g^2+4\lambda^2hc+4g^2-4+4hc+h^2c^2g^2-4hc g^2-h^2c^2}~.\nn
\eea
 where $c$ is an arbitrary constant.

Now at $\lambda=0$ we have the two parameter family of approximate
solutions
\bea\label{twolzero} g&=&1+\b\l^p+\dots\nn
h&=&\frac{2}{p}-\frac{2}{3p(2-p)\b}\l^{2-p}+\dots \eea for $0<p<2$.
This includes the exact solution when $p=1$ and $\b=1/{\sqrt 3}$.

For these solutions, near $\lambda=0$ we get
\bea
9m^2 \dd s^2\approx
4\frac{(p-c)^2}{p^2}[\dd\delta+\frac{1}{c-p}\sigma_3']^2
+\frac{1}{2\beta\lambda^p}\dd\lambda^2
+\frac{3}{p}[(\sigma_1')^2+(\sigma_2')^2]
+\frac{2\lambda^{2-p}}{(p-c)^2\beta}(
\sigma_3')^2~. \nonumber\\
\eea
We see that, for a given solution specified by $p,\beta$, this is
regular, with a two-sphere bolt, provided that the period of
$\gamma'$ is correlated with $c$. For example,
it will be useful to observe shortly that the period of $\gamma'$
can be taken to be $4\pi$ provided that $c=3p-4$ or $4-p$.

In order to mimic the PW solution, we would like to match these solutions
onto solutions with $h(\l_c)=0$ for some $\l_c$. Consider then the
one parameter family of
solutions\footnote{Note that there is also a two parameter family: \bea
g&=&(1+\l_c^2)^{1/2}-\frac{6}{5A}\e^{5/3}+\dots\nn h&=&A\e^{1/3}+\dots, \eea
but $h/g$, and hence the size of the two-sphere, diverges at $\epsilon=0$.}
\bea\label{onellc}
g&=&(1+\l_c^2)^{1/2}-\frac{2\l_c}{3(1+\l_c^2)^{1/2}}\e+\frac{3-\l_c^2}{27(1+\l_c^2)^{3/2}}\e^2+\dots\nn
h&=&\frac{
(1+\l_c^2)^{1/2}}{\l_c}\e+\frac{3-\l_c^2}{18\l_c^2(1+\l_c^2)^{1/2}}\e^2+\dots
\eea
with $\e=\l_c-\l$, which also includes the exact solution when $\l_c={\sqrt 3}$.

Now consider the behaviour of the metric for these solutions
\p{onellc} near $\lambda=\lambda_c$. We get, for all $\alpha$, \be
(9m^2) \dd s^2= 4[\dd\delta+D\sigma_3']^2
+\frac{3}{2\lambda_c\epsilon}\dd\epsilon^2
+\frac{3\epsilon}{2\lambda_c}[
(\sigma_1')^2+(\sigma_2')^2+(\sigma_3')^2] \ee with
$D\propto\epsilon$. This is regular provided that we take the
period of $\gamma'$ to be $4\pi$. We can now numerically integrate
these back to $\lambda=0$. The numerical analysis indicates that
they map onto a one parameter subset of the solutions
\p{twolzero}, with $3/2<h(0)\le 2$ {\it i.e.} $4/3>p\ge 1$. Thus we see
that if we choose $c=3p-4$ or $4-p$, then the base of the
fibration defined by $\partial_\delta$ is regular and has the
same topology as $\bCP^2$ (see the discussion above for the
Pilch--Warner solution). Furthermore, the numerical analysis
reveals that $A$ never vanishes and that $D$ only vanishes at
$\lambda_c$ in the way described above. We next note that
integrating $(1/2\pi)\diff (D\sigma_3')$ over the two-sphere bolt
at $\lambda=0$ gives $2D(0)=\pm 1/(p-2)$. Thus by choosing
the period of $\delta$ to be  $2\pi/(2-p)$ we deduce that the
topologies of this one-parameter family of solutions, generalising
the Pilch--Warner solution, are all $S^5$.

There are some problems with this family of solutions, however.
Firstly, from \p{esscond}, the spinor bilinear $S$ satisfies
$S=(\lambda/g)\me^{-2\ii\delta}$. For this to be well defined we
need the period of $\delta$ to be an integer times $\pi$. However,
for $4/3>p\ge 1$ this is only possible for the PW solution with
$p=1$. The second problem concerns the expression for the
flux~\p{fluxforspans}. For this to be globally well defined we should 
be able to write it in terms of globally defined one-forms
$\dd\delta$ and $\sigma_i'$. Note, however, $e^4-\ii
e^3\sim\sigma_2-\ii\sigma_1=\me^{\ii
  c\delta}(\sigma'_2-\ii\sigma_1')$, which requires $c$ to 
be an integer to be globally defined, and for $4/3>p\ge 1$ this is
again only possible for the PW solution which has $c=-1$ or 3. Note
that this phase in the expression for the complex three-form flux
cancels out in the energy momentum tensor and this is consistent
with the fact that the metric is globally defined on $S^5$. We
note that in appendix F, blindly ignoring these problems, we have
calculated the central charge of the putative dual conformal field
theories by determining the effective five-dimensional Newton's
constant.

Thus, to summarise, we conclude that these numerical
solutions for the ODE give rise to a regular metric on $S^5$ but
they do not give rise to a globally defined solution since the
three-form flux is not globally defined. Moreover, the Killing
spinors are also not globally defined. As discussed in
appendix F, the five-dimensional Newton's constant is, remarkably,
analytic for this family. In particular, the Newton's constant is
a monotonic decreasing function of $p$ for $1\leq p<4/3$. However,
the results of $a$-maximisation \cite{IW} in four-dimensional
superconformal field theories imply that the central charges are
always algebraic numbers. Indeed, in the current setting it is
natural to expect a quantisation condition on $p$ to come from
imposing well-definedness of the spinors and flux. As we have
shown, there are in fact no solutions to these conditions. It
seems possible that, nevertheless, there is some physical
interpretation of these solutions. Alternatively, we hope that by
slightly relaxing our assumptions new globally defined solutions
can be found.

\section{Conclusions}

The main result of this paper is a determination of the necessary
and sufficient conditions on supersymmetric solutions of type IIB
supergravity that can be dual to four-dimensional superconformal
field theories. The ten-dimensional metric is taken to be a warped
product of $AdS_5$ with a five-dimensional Riemannian metric and
we allowed for the most general fluxes consistent with $SO(4,2)$
symmetry. Excluding the well known $AdS_5\times X_5$ solutions
where $X_5$ is Sasaki--Einstein and only the self-dual five-form
is non-vanishing, we showed that the generic compact $M_5$ admits
a canonical local identity structure. We showed how supersymmetry
restricts the torsion of this structure and how it determines the
fluxes.

By imposing some additional restrictions, including that the
dilaton is constant and the axion vanishes, we reduced the
conditions to solving a second order non-linear ODE. We managed to
find an analytic solution of this ODE and showed that it
reproduces a solution found previously in \cite{pw}. It would be
nice to find the general solution of this ODE but our numerical
analysis is not  encouraging that there are further
globally defined solutions in this class. It would be interesting
to know if the local solutions that we found have a physical
interpretation. More generally, it may well be possible to find
new exact solutions by slightly relaxing some of our assumptions.

While this work was being completed, two papers appeared where new
classes of $AdS_5$ solutions of type IIB were discovered. In fact
the construction of these solutions was one of the original
motivations of this work. In \cite{Halmagyi:2005pn} numerical
evidence for a family of solutions interpolating between the PW
solution and the $AdS_5\times T^{1,1}$ solution were found. In
\cite{Lunin:2005jy} a powerful technique to generate new $AdS_5$
solutions from old ones, which describe the so-called
$\beta$-deformations of the original conformal field theory, was
presented. It would be interesting to see how these solutions fit
into the formalism presented here. It would be particulalry
interesting if the results of this paper could be used to find
$AdS_5$ solutions corresponding to exactly marginal deformations
more general than the $\beta$-deformations.

We only considered solutions preserving minimal $N=1$
supersymmetry. This includes geometries preserving $N=2$
supersymmetry as a special case, but it would be interesting to
determine the additional restrictions on the identity structure
that are imposed by $N=2$ supersymmetry.
Hopefully, these will be strong enough that further exact solutions can be found.
Recall
that in the context of D=11 supergravity, the analysis of
\cite{gmsw} covered $AdS_5$ geometries preserving $N=1$
supersymmetry. A refinement of this analysis was carried out in
\cite{Lin:2004nb}, where the additional conditions imposed by
$N=2$ supersymmetry were studied. It is interesting to note that a
double wick rotation of these geometries in \cite{Lin:2004nb} were
shown to be related to quite different physical phenomena, and
this may also be the case for the analogous type IIB supergravity
solutions.

Our analysis has focused on the local identity $G$-structure on
$M_5$, as this is most useful for obtaining explicit solutions. Of
course the category of families of solutions that can ultimately
be found in explicit form is presumably quite small. We also view
our work as providing the foundation for studying more general
aspects of conformal field theories with type IIB duals. For
example, it would be interesting to know what topological
restrictions supersymmetry imposes on $M_5$. To tackle this, one
could try to determine the global $G$-structure that $M_5$ admits.
A converse result of the form that $M_5$ satisfying certain
topological restrictions always admits a solution would be most
desirable. It would be also very interesting to see if
there is a generalisation of $Z$-minimisation \cite{MSY} (see also
\cite{Butti:2005vn,Tachikawa:2005tq,Barnes:2005bw}), a geometrical version of
$a$-maximisation \cite{IW} in the toric Sasaki--Einstein setting,
to the more general class of geometries analysed here.

\subsection*{Acknowledgments}
\noindent Most of this work was carried out whilst DM and JFS were
postdoctoral fellows at Imperial College, London. In particular DM
was funded by a Marie Curie Individual Fellowship under contract
number HPMF-CT-2002-01539, while JFS was supported by an EPSRC
mathematics fellowship. At present JFS is supported by NSF grants
DMS-0244464, DMS-0074329 and DMS-9803347.
DW is supported by a Royal Society University Research Fellowship and
thanks the Aspen Center for Physics for hospitality while this work
was being completed. JPG acknowledges partial support by
the National Science Foundation under Grant No. PHY99-0794 while
visiting the KITP, Santa Barbara.

\appendix

\section{Conventions and useful formulae for reduction}
\label{spinors}

The ten-dimensional metric has signature $(-,+,\dots,+)$. The
ten-dimensional gamma matrices $\Gamma^A$ satisfy
\begin{equation}
   \left[ \Gamma^A, \Gamma^B \right]_+ = 2 \eta^{AB}
\end{equation}
and generate the $D=10$ Clifford algebra $\Cliff(9,1)$, where
$A,B=0,1,\dots,10$ are frame indices. We define
$\Gamma_{11}\equiv\Gamma_0\Gamma_1\dots\Gamma_9$.

For the configurations that are a warped product of $AdS_5$ with
$M_5$, it is useful to decompose $\Cliff(9,1)$ by writing
\begin{equation}
\begin{aligned}
   \Gamma^{a} &= \rho^{a}\otimes 1 \otimes \sigma^3\\
   \Gamma^i &= 1\otimes \gamma^i \otimes \sigma^1
\end{aligned}
\end{equation}
where $a,b=0,1,\dots,4$ and $i,j=1,2,\dots,5$ are frame indices on
$AdS_5$ and $M_5$ respectively, and we have
\be
   \left[{\rho^a},{\rho^b}\right]_+ = 2\eta^{ab},\qquad
   \left[{\gamma^i},{\gamma^j}\right]_+
   =2\delta^{ij}\label{decomp}
\ee
with $\eta^{ab}=\diag(-1,1,1,1,1)$. The $\rho^a$ satisfy
$\rho_{01234}=\ii$ and generate $\Cliff(4,1)$, while the $\gamma^m$
satisfy $\gamma_{12345}=1$ and generate $\Cliff(5)$. In addition,
$\sigma^i$, $i=1,2,3$, are the Pauli matrices. We then have
\be
   \Gamma_{11}= 1\otimes 1 \otimes \sigma^2~.\ee

Let us work out a consistent set of conventions for the various
intertwiner operators in the relevant dimensions (see {\it e.g.}
\cite{sohnius}). The $A$-intertwiners operate as follows:
\bea A_{10} \Gamma^M A_{10}^{-1} & = & \Gamma^{M\dagger}\nn
A_{1,4} \rho^\mu A_{1,4}^{-1} & = -& \rho^{\mu\dagger}\nn A_{5}
\gamma^i A_{5}^{-1} & = & \gamma^{i\dagger} \eea
and can be chosen to be Hermitian: \bea A_\bullet^\dagger =
A_\bullet~. \eea The charge conjugation matrices, or
$C$-intertwiners, operate as follows:
\bea C_{10}^{-1} \Gamma^M C_{10} & = & -\Gamma^{M\trsp}\nn
C_{1,4}^{-1} \rho^\mu C_{1,4} & = & \rho^{\mu\trsp}\nn C_{5}^{-1}
\gamma^i C_{5} & = & \gamma^{i\trsp} \eea
and in the given dimensions are all antisymmetric:
\bea C_\bullet =  - C_\bullet^\trsp~. \eea
Finally, we have the following $D$-intertwiners
\bea \Gamma^{*A} =  \tilde{D}_{10}^{-1}\Gamma^{A}\tilde{D}_{10}
\quad
 \quad & & \tilde{D}_{10}\tilde{D}_{10}^*=1 \nn
\gamma^{*m} =  \tilde{D}_5^{-1}\gamma^{m}\tilde{D}_5 \quad \quad &
&  \tilde{D}_5\tilde{D}_5^*=-1 \nn \rho^{*a} =  -
D_{1,4}^{-1}\rho^{a}D_{1,4} \quad \quad & & D_{1,4}D_{1,4}^*=-1~.
\eea
Also recall that, by definition, $D_{10}  =  C_{10}
A_{10}^{\trsp}$ and that $\tilde D_{10}  =  \Gamma_{11} D_{10}$.
It turns out that one can take the following decompositions: \bea
A_{10} & = & 1 \otimes 1 \otimes \sigma^1\nn C_{10} & = & C_{1,4}
\otimes C_5 \otimes \sigma^2\nn
\tilde{D}_{10}&=&D_{1,4}\otimes\tilde{D}_5\otimes \sigma^1 \eea
with \bea A_{1,4}&=&1,\qquad C_{1,4}= D_{1,4}\nn A_5&=&1,\qquad
C_5=\tilde D_5~.
\eea
We now consider decomposing a $D=10$ Majorana--Weyl spinor $\e'$ as
$\epsilon'=\psi\otimes \chi\otimes \theta$. The chirality condition
in $D=10$ is \bea \Gamma_{11}\epsilon'=-\epsilon' \eea
which implies \bea \sigma^2\theta=-\theta~.\eea
Moreover, $\epsilon'^c=\tilde{D}_{10}\epsilon'^*$, which now reads
\bea \epsilon'^c = \psi^c\otimes \chi^c \otimes
\sigma^1\theta^*\eea
where \bea \psi^c&=&C_{1,4}\psi^*\nn \chi^c&=&C_5\chi^* \ee and we
note that $\psi^{cc}=-\psi$ and $\chi^{cc}=-\chi$. To impose the
Majorana condition in $D=10$, $\epsilon'^c=\epsilon'$, we take \be
\theta=\sigma^1\theta^* \ee which we note is consistent with the
chirality condition already imposed on $\theta$.

We now want to construct the most general spinor ansatz that is
consistent with minimal supersymmetry in $AdS_5$. Since type IIB
supersymmetry is parametrised by two $D=10$ Majorana--Weyl spinors,
$\e_i$, we take
\be \epsilon_i = \psi\otimes \chi_i\otimes \theta+
\psi^c\otimes\chi_i^c\otimes \theta\ee
where we assume that the spinor $\psi$ satisfies
\bea \nabla_{\mu}\psi & = & \frac{1}{2}m\rho_{\mu}\psi \eea
to ensure that supersymmetry is preserved on $AdS_5$. Notice that
$\psi^c$ then satisfies this equation with $m\mapsto -m$. Notice
also that each spinor has 16 real components, realised as the real
part of 4 complex times 4 complex components. We may then
complexify
\be
\epsilon \equiv \epsilon_1+\ii\epsilon_2 \equiv \psi\otimes
\xi_1\otimes\theta + \psi^c\otimes \xi_2^c\otimes \theta
\ee
where $\xi_1=\chi_1+\ii\chi_2$, $\xi_2^c=\chi_1^c+\ii\chi_2^c$. Then
\be \epsilon^c = \psi\otimes \xi_2\otimes \theta+\psi^c\otimes
\xi_1^c\otimes \theta~.\ee
In fact, to derive \p{sone}--\p{ssix}, we rescaled by a convenient
power of the warp factor. Indeed the ansatz we used is:
\bea\label{spinans} \epsilon \equiv \psi\otimes
\me^{\Delta/2}\xi_1\otimes \theta + \psi^c\otimes
\me^{\Delta/2}\xi_2^c\otimes \theta~. \eea
After substituting this into the Killing spinor equations
\p{10d-susy}, one finds a number of equations.

In order to analyse the Killing spinor equations, we shall also
need the following result. Suppose we have two complex vector
spaces $V$ and $W$, such that $V$ comes equipped with an
anti-unitary operation $c$, mapping $v\in V$ to $v^c\in V$, with
$(av)^c=a^*v^c$ $\forall a\in\mathbb{C}$, which also squares to
$-1$: $v^{cc}=-v$. Then for non-zero $v\in V$, we have $v\otimes
w_1 + v^c\otimes w_2=0$ implies that $w_1=w_2=0$. To see this,
first note that $v\otimes w_1 + v^c\otimes w_2=0$ implies that
either $w_1=w_2=0$, or else $v=av^c$, $w_1=w_2/a$ for some
$a\in\mathbb{C}^*$. Suppose the latter case holds. Taking the
conjugate of $v-av^c=0$ gives $v^c+a^*v=0$. These two equations
imply $1+|a|^2=0$, which is impossible. Hence the result. Using
this algebraic lemma one can show that the equations are then a
sum of two terms, each of which is separately zero.

Much of our analysis of the supersymmetry conditions
\p{sone}--\p{ssix} come from analysing bilinears that can be
constructed from $\xi_i$. Note that there are two kinds of
bilinears that can be constructed \bea \bar\chi \gamma_{(n)}\xi &
= & \chi^\dagger \gamma_{(n)}\xi\nn \bar\chi^c \gamma_{(n)}\xi & =
& \chi^\trsp C^{-1} \gamma_{(n)}\xi \eea where we have used
$A_5=1$, defined $C\equiv C_5 = \tilde D_5$ and $\gamma_{(n)}$ is
the antisymmetrised product of $n$ gamma matrices. For convenience
we record once again, for reference: \bea C^* & = & -C^{-1}\nn
C^\trsp & = & - C \eea and \bea \gamma^i  & = &
\gamma^{i\dagger}\nn C^{-1} \gamma^i C & = & \gamma^{i\trsp}~.
\eea
Finally we note that the Fierz identity for $\Cliff(5)$ reads: \be
\bar\xi_1\xi_2\bar\xi_3\xi_4= \frac{1}{4}\bar\xi_1\xi_4
\bar\xi_3\xi_2 +\frac{1}{4}\bar\xi_1\gamma_m\xi_4
\bar\xi_3\gamma^m\xi_2 -\frac{1}{8}\bar\xi_1\gamma_{mn}\xi_4
\bar\xi_3\gamma^{mn}\xi_2~. \ee

\section{Algebraic conditions satisfied by the bilinears}
\label{basis}

There are a number of algebraic conditions satisfied by the
various bilinears that we use in the main text that can be derived
using Fierz identities. However, we find it most useful to
construct them using a convenient basis of $\gamma$-matrices of
$\Cliff(5)$. Specifically, we start by taking
\begin{equation}
\begin{aligned}
   \gamma^1 &=
      \begin{pmatrix} 1 & 0 \\ 0 & -1 \end{pmatrix} \otimes \id \\
   \gamma^2 &=
      \begin{pmatrix} 0 & 1 \\ 1 & 0 \end{pmatrix} \otimes \id \\
   \gamma^a &=
      \begin{pmatrix} 0 & -1 \\ 1 & 0 \end{pmatrix} \otimes \tau^a
\end{aligned}
\end{equation}
where $\tau^a=-\ii\sigma^a$ and $\sigma^a$ are the Pauli matrices.
The intertwiner $\gamma^{*m}=\tilde{D}^{-1}\gamma^m\tilde{D}$,
used in the definition $\xi^c=\tilde{D}\xi^*$, is given by
$D=\id\otimes -\ii\sigma^2$. The corresponding basis of one-forms
are labelled $e^i$.

We write the two spinors as $\xi_i=s_i\otimes\theta_i$. We demand
that the two vectors, ${K}_4$ and ${K}_5$ defined in \p{bilins}
are chosen to lie in the ($e^1$--$e^2$)-plane. This requires
$\bar\xi_i\gamma^a\xi_i=0$ and constrains the $s_i$. If in
addition we require ${K}_5$ to be parallel to $e^1$ we find
\begin{equation}
   s_1 = \sqrt{2}\begin{pmatrix} \cos\theta\cos\phi \\
      -\sin\theta\sin\phi \end{pmatrix} \qquad
   s_2 = \sqrt{2}\begin{pmatrix} \sin\theta\cos\phi \\
      \cos\theta\sin\phi \end{pmatrix} \qquad
\end{equation}
where, without loss of generality, we have
$\bar{\theta}_i\theta_i=1$. Note that we have imposed \p{eq:dA}:
\begin{equation}
   \bar{\xi}_1\xi_1 + \bar{\xi}_2\xi_2 = 2~.
\end{equation}

One then finds that the $\theta$ and $\phi$ functions are related
to the scalar bilinears $\sin\zeta$, $Z$ and $S$ defined in
\p{bilins}, by
\begin{equation}
\begin{aligned}
   \sin\zeta &= \cos2\theta\cos2\phi \\
   Z &= \sin2\theta\cos2\phi\ \bar{\theta}_2\theta_1 \\
   S &= \sin2\theta\cos2\phi\ \bar{\theta}^c_2\theta_1~.
\end{aligned}
\end{equation}
The vectors defined in \p{bilins} are given by
\begin{equation}
\begin{aligned}
   K_5 &= (\cos2\phi) e^1 \\
   K_4 &= (\cos2\theta) e^1 - (\sin2\theta\sin2\phi) e^2 \\
   K_3 &= (\sin2\theta\ \bar{\theta}_2\theta_1) e^1
      + (\cos2\theta\sin2\phi\ \bar{\theta}_2\theta_1) e^2
      + (\sin2\phi) \bar{\theta}_2\tau_a\theta_1\, e^a \\
   K &= (\sin2\theta\ \bar{\theta}^c_1\theta_2) e^1
      + (\cos2\theta\sin2\phi\ \bar{\theta}^c_1\theta_2) e^2
      - (\sin2\phi) \bar{\theta}^c_1\tau_a\theta_2\, e^a~.
\end{aligned}
\end{equation}
It is similarly straightforward to write out the two-forms. In
particular, we find
\begin{multline}
   W = Z (\csc2\theta\tan2\phi) \,e^{12}
      + \left(\cos2\theta\sin2\phi \,e^1 - \sin2\theta \,e^2\right)
         \wedge \bar{\theta}_2\tau_a\theta_1 \, e^a \\
      + \frac{1}{2}(\sin2\theta\cos2\phi) \epsilon_{abc}
         \bar{\theta}_2\tau^c\theta_1 \,e^{ab}~.
\end{multline}
We then find that $\re[\ii Z^*W]=0$ implies $Z=0$ as claimed in the
text.

We now put $Z=0$ by setting $\bar{\theta}_1\theta_2=0$. Choosing
$K_3$ to suitably lie just within the ($e^3$--$e^4$)-plane, we can
choose
\begin{equation}
   \theta_1 = \begin{pmatrix} \me^{\ii\alpha} \\
       0\end{pmatrix} \qquad
   \theta_2 = \begin{pmatrix} 0 \\
      \me^{\ii\alpha}\end{pmatrix} \qquad
\end{equation}
and hence
\begin{equation}
\begin{aligned}
   K_5 &= (\cos2\phi) e^1 \\
   K_4 &= (\cos2\theta) e^1 - (\sin2\theta\sin2\phi) e^2 \\
   K_3 &= (\sin2\phi)(e^4 - \ii e^3) \\
   \me^{-2\ii\alpha} K &=
      (\sin2\theta) e^1 + (\cos2\theta\sin2\phi) e^2
         - \ii (\sin2\phi) e^5
\end{aligned}
\end{equation}
where $\bar{\theta}^c_1\theta_2=\me^{2\ii\alpha}$, which is the
phase of $S$, so that the scalars are now given by
\begin{equation}
\begin{aligned}
   \sin\zeta &= \cos2\theta\cos2\phi \\
   S &= - \sin2\theta\cos2\phi\, \me^{2\ii\alpha}~.
\end{aligned}
\end{equation}
Similarly, the two-forms are
\begin{equation}
\begin{aligned}
   U &= -\sin2\theta\sin2\phi \,e^{15} - \cos2\theta \,e^{25}
      + \cos2\theta\cos2\phi \,e^{34} \\
   V &= -\cos2\phi \,e^{25} + e^{34} \\
   W &= \left( \cos2\theta\sin2\phi \,e^1 - \sin2\theta \,e^2
      + \ii\sin2\theta\cos2\phi \,e^5\right) \wedge (e^4 - \ii e^3)
\end{aligned}
\end{equation}
and this leads to a quick derivation of \p{UVWexp}.

\section{Absence of solutions with $\xi_i$ linearly dependent and $\xi_i\ne 0$}
\label{absol}

Let us consider the possibility \be
\xi_2=u\xi_1+v\xi_1^c \ee for some functions $u,v$, which defines
a local $SU(2)$ structure in five dimensions. We will use the
conditions \p{UVrel} and \p{linrel} that can be derived directly
from Fierz relations and in particular do not rely on any aspects
of the identity structure that we considered in the text. We also
use the differential conditions \p{vK}--\p{v5}. Recall from section
3.2 that we  can then deduce \p{fxf} and $\re [\ii Z^* W]  =  0$.

A calculation shows that $\re [\ii Z^* W]  =  0$ implies that \be
2|u|^2 \bar\xi_1\gamma_{(2)}\xi_1
+uv^*\bar\xi_1^c\gamma_{(2)}\xi_1
+u^*v\bar\xi_1\gamma_{(2)}\xi_1^c=0~. \ee In the special case that
$v=0$ we deduce, for non-trivial $\xi_1$, that $u=0$ and we return
to the Sasaki--Einstein case analysed just before the start of
section 3.1. We continue, therefore, with $v\ne0$. To proceed we
derive the following expression for $V$: \be\label{veeexpap}
V=-\frac{1+|u|^2+|v|^2}{2}\ii \bar\xi_1\gamma_{(2)}\xi_1~. \ee
Next observe that $K=-S K_5$ and using \p{s2} we get \be -3\ii
mK_5=\diff[\mathrm{ln}(\me^{4\Delta}S)] \ee and hence $\diff
K_5=0$. We also have $K_4=\sin\zeta K_5$ and then using \p{fixf}
we get \be \me^{4\Delta}K_4=\frac{f}{4m}K_5~. \ee {} From \p{v4}
we then conclude that $V=0$. However, from \p{veeexpap} we see
that this is not possible unless $\xi_1=0$.

\section{Integrability of IIB supersymmetry conditions}
\label{integrab}

Writing the variation of the gravitino appearing in \p{10d-susy}
as $\delta\psi_M={\cal D}_M\e$, we calculate that \bea {\cal
D}_{[M} {\cal D}_{N]}\e=I^{(1)}_{MN}\e+ I^{(2)}_{MN}\e^c \eea
where \bea
I^{(1)}_{MN}&=&\frac{1}{8}R_{MNS_1S_2}\Gamma^{S_1S_2}-\frac{1}{2}P_{[M}P^*_{N]}
+\frac{\ii}{192}D_{[M}F_{N]S_1S_2S_3S_4}\Gamma^{S_1S_2S_3S_4}\nn
&&+\frac{1}{768}\big(F_{MNS_1}{}^{R_1R_2}F_{S_2S_3S_4R_1R_2}\Gamma^{S_1S_2S_3S_4}
-2F_{[M|S_1|}{}^{R_1R_2R_3}F_{N]S_2R_1R_2R_3}\Gamma^{S_1S_2}\big)\nn
&&+\frac{1}{9216}\big(
-\Gamma_{MN}{}^{S_1S_2S_3S_4S_5S_6}G_{S_1S_2S_3}G^*_{S_4S_5S_6}
-9\Gamma_{MN}{}^{S_1S_2S_3S_4}G_{S_1S_2}{}^RG^*_{S_3S_4R}\nn
&&~~~~~~~~~~+12\Gamma_{[M}{}^{S_1S_2S_3S_4S_5}G_{N]S_1S_2}G^*_{S_3S_4S_5}
-6\Gamma_{[M}{}^{S_1S_2S_3S_4S_5}G_{|S_1S_2S_3|}G^*_{N]S_4S_5}\nn
&&~~~~~~~~~~+18\Gamma_{MN}{}^{S_1S_2}G_{S_1}{}^{R_1R_2}G^*_{S_2R_1R_2}
+36\Gamma_{[M}{}^{S_1S_2S_3}G_{|S_1S_2|}{}^RG^*_{N]S_3R}\nn
&&~~~~~~~~~~+72\Gamma_{[M}{}^{S_1S_2S_3}G_{N]S_1}{}^RG^*_{S_2S_3R}
+72\Gamma^{S_1S_2S_3S_4}G_{[M|S_1S_2|}G^*_{N]S_3S_4}\nn
&&~~~~~~~~~~-18\Gamma^{S_1S_2S_3S_4}G_{S_1S_2S_3}G^*_{MNS_4}
+18\Gamma^{S_1S_2S_3S_4}G_{MNS_1}G^*_{S_2S_3S_4}\nn
&&~~~~~~~~~~+6\Gamma_{MN}G^{R_1R_2R_3}G^*_{R_1R_2R_3}
-72\Gamma_{[M}{}^{S}G_{N]}{}^{R_1R_2}G^*_{SR_1R_2}\nn
&&~~~~~~~~~~+36\Gamma_{[M}{}^{S}G_{|S|}{}^{R_1R_2}G^*_{N]R_1R_2}
-288\Gamma^{S_1S_2}G_{[M |S_1|}{}^{R}G^*_{N]S_2R}\nn
&&~~~~~~~~~~-54\Gamma^{S_1S_2}G_{S_1S_2}{}^RG^*_{MNR}
+54\Gamma^{S_1S_2}G_{MN}{}^RG^*_{S_1S_2R}\nn
&&~~~~~~~~~~-144G_{[M}{}^{R_1R_2}G^*_{N]R_1R_2}\big) \eea and \bea
I^{(2)}_{MN}&=&
\frac{1}{96}\big(\Gamma_{[M}{}^{S_1S_2S_3}D_{N]}G_{S_1S_2S_3}
+9\Gamma^{S_1S_2}D_{[M}G_{N]S_1S_2}\big)\nn
&&-\frac{\ii}{1536}\big(
3G_{MN}{}^RF_{RS_1S_2S_3S_4}\Gamma^{S_1S_2S_3S_4}
+6G_{[M}{}^{R_1R_2}F_{R_1R_2S_1S_2S_3}\Gamma_{N]}{}^{S_1S_2S_3}\nn
&&~~~~~~~-6G_{S_1}{}^{R_1R_2}F_{[M|R_1R_2S_2S_3|}\Gamma_{N]}{}^{S_1S_2S_3}
-6G_{S_1S_2}{}^{R}F_{MNRS_3S_4}\Gamma^{S_1S_2S_3S_4}\nn
&&~~~~~~~-12G_{[M}{}^{R_1R_2}F_{N]R_1R_2S_1S_2}\Gamma^{S_1S_2}
+4G^{R_1R_2R_3}F_{[M|R_1R_2R_3S|}\Gamma_{N]}{}^{S}\nn
&&~~~~~~~+4G^{R_1R_2R_3}F_{MNR_1R_2R_3}\big) \eea and we have used
the self-duality of $F$. Note that this result is consistent with
that in \cite{Papadopoulos:2003jk}.

Setting this to zero, contracting with $8\Gamma^N$ we deduce that
\bea\label{gravint} &&2\big[-R_{MS} +P_M P_S^*+P_SP^*_M +
\frac{1}{96}F_{M}{}^{R_1R_2R_3R_4}F_{SR_1R_2R_3R_4}\nn
&&~~~~~~~~~+ \frac{1}{8}\big(G_M{}^{R_1R_2}G^*_{SR_1R_2} +
G_S{}^{R_1R_2}G^*_{MR_1R_2}
         - \frac{1}{6}g_{MS}G^{R_1R_2R_3}G^*_{R_1R_2R_3}\big)\big]\Gamma^S\e\nn
&&-\frac{\ii}{48}*_{10}[\diff F-\frac{\ii}{2}G\wedge
G^*]_{S_1S_2S_3S_4}\Gamma_M{}^{S_1S_2S_3S_4}\e
+\frac{\ii}{12}*_{10}[\diff F-\frac{\ii}{2}G\wedge
G^*]_{MS_1S_2S_3}\Gamma^{S_1S_2S_3}\e\nn
&&-\frac{1}{96}(DG+P\wedge
G^*)_{S_1S_2S_3S_4}\Gamma_M{}^{S_1S_2S_3S_4}\e^c
+\frac{1}{8}(DG+P\wedge G^*)_{MS_1S_2S_3}\Gamma^{S_1S_2S_3}\e^c\nn
&&-\frac{1}{8}(D_RG^R{}_{S_1S_2}+\frac{\ii}{6}G_{R_1R_2R_3}F^{R_1R_2R_3}{}_{S_1S_2}
-P^RG^*_{S_1S_2R})\Gamma_M{}^{S_1S_2}\e^c\nn
&&+\frac{3}{4}(D_RG^R{}_{MS}+\frac{\ii}{6}G_{R_1R_2R_3}F^{R_1R_2R_3}{}_{MS}
-P^RG^*_{MSR})\Gamma^S\e^c\nn
&&~~~~~~=-\frac{\ii}{24}G^*_{S_1S_2S_3}\Gamma_M{}^{S_1S_2S_3}\delta\lambda
+\frac{3\ii}{8}G^*_{MS_1S_2}\Gamma^{S_1S_2}\delta\lambda
+4\ii P_M\delta\lambda^*~. \eea

Similarly, again using the self-duality of $F$, a calculation
reveals that \bea\label{dilint} &&\ii
(D_MP^M+\frac{1}{24}G^{M_1M_2M_3}G_{M_1M_2M_3})\e^c +\ii
D_{[S_1}P_{S_2]}\Gamma^{S_1S_2}\e^c\nn
&&+\frac{\ii}{24}(D_{[S_1}G_{S_2S_3S_4]}+P_{[S_1}
G^*_{S_2S_3S_4}])\Gamma^{S_1S_2S_3S_4}\e\nn
&&+\frac{\ii}{8}(D_MG^M{}_{S_1S_2}-P_MG^{*M}{}_{S_1S_2}
+\frac{\ii}{6}F_{S_1S_2M_1M_2M_3}G^{M_1M_2M_3})\Gamma^{S_1S_2}\e\nn
= &&\Gamma^S D_S \delta \lambda
-\frac{\ii}{960}\Gamma^{S_1S_2S_3S_4S_5}
F_{S_1S_2S_3S_4S_5}\delta\lambda-\ii\Gamma^M\Gamma^S P_S\delta
\psi_M^*\nn
&&-\frac{\ii}{24}\Gamma^M\Gamma^{S_1S_2S_3}G_{S_1S_2S_3}\delta\psi_M~.
\eea

Suppose we have a supersymmetric configuration satisfying $\delta
\psi_M=\delta\lambda=0$. If we demand that it also satisfies the
equation of motion and the Bianchi identity for $G$, the Bianchi
identity for $P$ and the Bianchi identity for the self-dual
five-form $F$, then we conclude from \p{dilint} that the equation
of motion for $P$ is automatically satisfied and from \p{gravint}
that \be\label{condein} E_{MS}\Gamma^S\e=0 \ee where $E_{MS}=0$ is
equivalent to Einstein's equations. Now the vector bilinear,
$K^M\equiv \bar \e \Gamma^M\e$ that can be constructed from a
spinor of $Spin(9,1)$ is null. We can use it to set up a frame
\be\label{frameint} \dd s^2=2e^+ e^- + e^a e^a \ee for $a=1,\dots,
9$ with $K$, as a one-form, equal to $e^+$. Following the argument
of section 2.3 of \cite{Gauntlett:2002fz}, we conclude that
\p{condein} implies that the only component of $E_{MS}$ that is
not automatically zero is $E_{++}$, which thus is the only extra
condition that needs to be imposed in order to get a full
supersymmetric solution to the equations of motion.

For the class of solutions considered in the text, we have
$E_{++}=0$. To see this, we first note that the spinor ansatz
\p{spinans} implies that the vector $K^M$ only has non-vanishing
components in the $AdS_5$ directions. Next we observe that the
Ricci tensor of the ten-dimensional metric has components in the
$AdS_5$ directions given by \be R_{\mu\nu}=-\bar
g_{\mu\nu}(4m^2+8(\bar\nabla \Delta)^2+ \bar\nabla^2\Delta) \ee
where $\bar g$ is the metric on $AdS_5$. In addition the right
hand side of the Einstein equation in \p{10d-susy} is also
proportional to $\bar g$. Since $\bar g_{++}=0$, in the frame
\p{frameint}, we conclude that $E_{++}=0$ is trivially satisfied.

\section{Central charges}

Recall that the central charge of the conformal field theory dual
is determined by the five-dimensional Newton's
constant~\cite{Henningson:1998gx}. The type IIB action takes the form
\be 
   S = \frac{1}{2\kappa^2_{10}} 
      \int_{M} \sqrt{G}\left[R(G)+\dots\right]~,
\ee
where $G$ is the ten-dimensional metric. Assuming $M=M_{4,1}\times
M_5$ and using our warped-product ansatz~\eqref{metansatz}, we get 
\be
   S= \frac{1}{2\kappa^2_{10}} \int_{M_5}\sqrt{g_5}\,\me^{8\Delta}
      \int_{M_{4,1}}\sqrt{g_E}\left[R(g_E)+12m^2\dots\right]~, 
\ee
where $g_E$ and $g_5$ are the metrics on $M_{4,1}$ and $M_5$
respectively. The $m^2$ term in the integrand appears since our
$AdS_5$ metric on $M_{4,1}$ is normalised so that
$\Ric(g_E)=-4m^2g_E$. We will also need the quantisation condition on
the five-form which reads  
\be \int_{M_5} F-\frac{\ii}{2} A\wedge G^* = c_0N_{\mathrm{D3}} 
\ee 
where $N_{\mathrm{D3}}$ is the number of $D3$-branes and  $c_0$ is a
constant the precise form of which we will not need.

Let us first consider the simplest case of where $M_5$ is
Sasaki--Einstein. We then have $\me^{8\Delta}=(f/4m)^2$, and
the quantisation condition gives $f=c_0N_{\mathrm{D3}}m^5/\vol'(M_5)$
where $\mathrm{vol}'(M_5)$ is the volume of the Sasaki--Einstein
metric normalised so that $\Ric(g_5)=4g_5$. In particular
$\vol'(S^5)=\pi^3$. Thus the type IIB dimensionally reduced action
reads  
\be
   S=\frac{1}{16\pi G_5^{\SE}} \int_{M_{4,1}}\sqrt{g_E}
      \left[R(g_E)+12m^2\dots\right]~,
\ee
where
\be
  G_5^{\SE}=\frac{2\kappa^2_{10}\vol'(M_5)}
    {\pi m^3c_0^2N_{\mathrm{D3}}^2}~.
\ee

Next we consider the special class of solutions that we discussed
in section 5. We first observe 
\be\label{whatwefirst} 
   \int_{M_5}\sqrt{g_5}\,\me^{8\Delta}
      = -\frac{2f^2}{16^2\times 27 m^7}\int_{M_5}
         \,\de_\lambda(h^2)\,\diff\lambda\,\diff\psi\,\diff y\, 
            \sigma_1\wedge\sigma_2
\ee 
noting the remarkable fact that the integrand can be trivially
integrated. Equally remarkable is the fact that the 
quantisation condition on the flux also takes a simple form. To
carry out the integral, we first observe that a two-form potential
for the three-form flux $G$ is given by 
\be 
\label{Aform}
   A = \frac{1}{3m^2}\left(\frac{f}{4m}\right)^{1/2}
      \frac{h^{1/2}(g^2-1-\lambda^2)^{1/2}}{\sqrt{6}\lambda g^2}
      \left(-\ii\diff\lambda+g\lambda \diff\psi\right)
      \wedge\left(\sigma_2-\ii\sigma_1\right)~. 
\ee
It is then straightforward to deduce that 
\be
   f^{-1} = \frac{1}{27\times 8m^5 c_0N_{\mathrm{D3}}}
      \int_{M_5}\de_\lambda(h^2/g^2)\,\diff\lambda\,\diff\psi\,\diff y\,
       \sigma_1\wedge\sigma_2~. 
\ee 
Substituting this into~\p{whatwefirst} gives an analytic expression
for the Newton's constant for this class of solutions. 

We finally focus on the local solutions that we found numerically that
include the PW solution as a special case. In particular $0\le
\lambda\le \lambda_c$ with $h(\lambda_c)=0$ and $g(0)=1$. In this case
we get 
\be\label{troublesome}
   G_5^{(\mathrm{sec}5)} = 
      -\frac{\kappa^2_{10}h^2(0)}{4\times 27\pi m^3c_0^2 N^2_{D3}}
      \int \diff\psi\,\diff y\,\sigma_1\wedge\sigma_2~. 
\ee
For the numerical solutions, $\sigma_1^2+\sigma_2^2$ gives the round
metric on $S^2$ so $\int\sigma_1\wedge\sigma_2=4\pi$, while $\diff
\psi\wedge \diff  y=2\diff \delta\wedge \diff \gamma'$ and so when
integrated contributes a factor of $16\pi^2/(2-p)$. Recall that 
$h(0)=2/p$ and that the value of $p$ is determined by $\lambda_c$
which specifies the numerical solution and that for the PW solution we
have $p=1$. 

As a check on these formulae we can calculate the ratio of the central
charges of the theories dual to $AdS_5\times S^5$ and the PW
solution. We get
\be
\frac{G_5^{\mathrm{PW}}}{G_5^{S^5}}=\frac{32}{27}
\ee
in agreement with \cite{kpw}.

More generally, the expression \p{troublesome} shows that the
ratio of central charges of two solutions in our new family of
local solutions depends on the ratio of their values of
$1/p^2(2-p)$. As this is not constant it indicates that the local
solutions  could not possibly represent exactly marginal
deformations of the PW solution.  Furthermore, if the local
solutions were to somehow make physical sense, some restrictions
on $p$ would have to be imposed to ensure an algebraic central
charge as implied by the general results on $a$-maximisation
\cite{IW}.


\end{document}